\documentclass{stavanger}

\usepackage[utf8]{inputenc}
\usepackage{amssymb,amsmath,amsfonts,amsthm}
\usepackage{hyperref}
\usepackage{natbib}
\usepackage{graphicx}
\usepackage{tikz}
\usepackage{cleveref}
\usepackage{mathtools}
\usepackage{placeins}
\usepackage{xcolor}
\usepackage[export]{adjustbox} % Aligning figures left and right
\usepackage{courier}
\usepackage{bm}

\startlocaldefs

\endlocaldefs

\begin{document}

%%%%%%%%%%%%%%%%%%%%%%%%%%%%%%%%%%%%%%%%%
%%                                     									       %%
%%  Start of the title page information section						       %%
%%                                     									       %%
%%%%%%%%%%%%%%%%%%%%%%%%%%%%%%%%%%%%%%%%%

\begin{frontmatter}

\begin{fmbox}

%%%%%%%%%%%%%%%%%%%%%%%%%%%%%%%%%%%%%%%%%
%%                                     									       %%
%%  Set the header on the first page, default is "Research Article"		       %%
%%                                     									       %%
%%  Define which division of the Institute of Mathematics and Physics	       %%
%%  this work is associated with								       %%
%%                                     									       %%
%%        PM:	Pure Mathematics                         					       %%
%%        ST:	Mathematical Statistics                     					       %%
%%        MP:	Mathematical Physics                        					       %%
%%        MS:	Material Physics                         					       %%
%%        FP:	Theoretical Physics							       %%
%%                                     									       %%
%%%%%%%%%%%%%%%%%%%%%%%%%%%%%%%%%%%%%%%%%

\dochead{Methods Review - Preprint}{FP}

%%%%%%%%%%%%%%%%%%%%%%%%%%%%%%%%%%%%%%%%%
%%                                     									       %%
%%  Title of the manuscript									       %%
%%                                     									       %%
%%%%%%%%%%%%%%%%%%%%%%%%%%%%%%%%%%%%%%%%%

\title{Bayesian inference of real-time dynamics from lattice QCD}

%%%%%%%%%%%%%%%%%%%%%%%%%%%%%%%%%%%%%%%%%
%%                                     									       %%
%%  Author information										       %%
%%                                     									       %%
%%%%%%%%%%%%%%%%%%%%%%%%%%%%%%%%%%%%%%%%%

\author[
   addressref={aff1},                   	  % id's of addresses
   %corref={aff1},                     		  % id of corresponding address, if any
   %noteref={n1},                        		  % id's of article notes, if any
   email={alexander.rothkopf@uis.no}   		  % email address
]{\inits{AR}\fnm{Alexander} \snm{Rothkopf}}

%%%%%%%%%%%%%%%%%%%%%%%%%%%%%%%%%%%%%%%%%
%%                                     									       %%
%%  Affiliations of the authors									       %%
%%                                     									       %%
%%%%%%%%%%%%%%%%%%%%%%%%%%%%%%%%%%%%%%%%%

\address[id=aff1]{%                          			 % unique id
  \orgname{Faculty of Science and Technology}, 	 % faculty
  \street{University of Stavanger},                     		 % university
  \postcode{4021}                               			 % post or zip code
  \city{Stavanger},                              				 % city
  \cny{Norway}                                   				 % country
}

%%%%%%%%%%%%%%%%%%%%%%%%%%%%%%%%%%%%%%%%%
%%                                     									       %%
%%  Additional notes on the authors								       %%
%%                                     									       %%
%%%%%%%%%%%%%%%%%%%%%%%%%%%%%%%%%%%%%%%%%

%\begin{artnotes}
%\note[id=n1]{Only contributor} % note, connected to author
%\end{artnotes}

\end{fmbox}% comment this for two column layout

%%%%%%%%%%%%%%%%%%%%%%%%%%%%%%%%%%%%%%%%%
%%                                     									       %%
%%  Abstract of the manuscript									       %%
%%                                     									       %%
%%%%%%%%%%%%%%%%%%%%%%%%%%%%%%%%%%%%%%%%%

\begin{abstractbox}

\begin{abstract} % abstract
The computation of dynamical properties of nuclear matter, ranging from parton distribution functions of nucleons and nuclei to transport properties in the quark-gluon plasma, constitutes a central goal of modern theoretical physics. This real-time physics often defies a perturbative treatment and the most successful strategy so far is to deploy lattice QCD simulations. These numerical computations are based on Monte-Carlo sampling and formulated in an artificial Euclidean time. Real-time physics is most conveniently formulated in terms of spectral functions, which are hidden in lattice QCD behind an ill-posed inverse problem. I will discuss the current methods state-of-the art in the extraction of spectral functions from lattice QCD simulations, based on Bayesian inference and emphasize the importance of prior domain knowledge, vital to regularizing the otherwise ill-posed extraction task. With Bayesian inference allowing us to make explicit the uncertainty in both observations and in our prior knowledge, a systematic estimation of the total uncertainties in the extracted spectral functions is nowadays possible. Two implementations of the Bayesian Reconstruction (BR) method for spectral function extraction, one for MAP point estimates and one based on an open access Monte-Carlo sampler are provided.
I will briefly touch on the use of machine learning for spectral function reconstruction and discuss some new insight it has brought to the Bayesian community.
\end{abstract}

%%%%%%%%%%%%%%%%%%%%%%%%%%%%%%%%%%%%%%%%%
%%                                     									       %%
%%  Keywords for the article, each one in its separate \kwd{}			       %%
%%                                     									       %%
%%%%%%%%%%%%%%%%%%%%%%%%%%%%%%%%%%%%%%%%%


\end{abstractbox}

%\end{fmbox}% uncomment this for twcolumn layout

\end{frontmatter}

%%%%%%%%%%%%%%%%%%%%%%%%%%%%%%%%%%%%%%%%%
%%                                     									       %%
%%  Main text starts here										       %%
%%                                     									       %%
%%%%%%%%%%%%%%%%%%%%%%%%%%%%%%%%%%%%%%%%%

\section{Introduction}
\label{sec:introduction}

\subsection{The Physics Challenge}
After a successful decade of studying the static properties of the strong interactions, such as their phase diagram (for reviews see e.g. \cite{Guenther:2020jwe,Fukushima:2010bq}) and equation of state (for recent studies see e.g. \cite{Borsanyi:2022soo,Bazavov:2017dus,Borsanyi:2022qlh}) through relativistic heavy-ion collisions (for an overview see e.g. \cite{Busza:2018rrf}) and more recently through the multi-messenger observations of colliding neutron stars (for a review see e.g. \cite{Kojo:2020krb}), high energy nuclear physics sets out to make decisive progress in the understanding of real-time dynamics of quarks and gluons in the coming years. 

The past heavy-ion collision campaigns at collider facilities such as RHIC at Brookhaven National Laboratory (BNL) and the LHC at the European Center for Nuclear Physics (CERN) provided conclusive evidence for the existence of a distinct high-temperature state of nuclear matter, the quark-gluon-plasma (for a review see e.g. \cite{Pasechnik:2016wkt}). At the same time, theory, by use of high-performance computing, predicted the thermodynamic properties, such as the equation of state \cite{Bazavov:2017dsy,Borsanyi:2016ksw,HotQCD:2014kol,Burger:2014xga,Borsanyi:2013bia} of hot nuclear matter from first principles. When data and theory were put to the test in the form of phenomenological models based on relativistic hydrodynamics, excellent agreement was observed (for a review see e.g. \cite{Jaiswal:2016hex}). 

Similarly past $e^-$+$p$ collider experiments at HERA (DESY) revealed (for a review see \cite{Klein:2008di}) that the properties of nucleons can only be understood when in addition to the three valence quarks of the eponymous quark-model also the virtual excitations of quarks and gluons are taken into account. In particular the emergent phenomenon of asymptotic freedom manifests itself clearly in their data, as the coupling between quarks and gluons becomes weaker with increasing momentum exchange in a collision (for the current state-of-the art see e.g \cite{dEnterria:2022hzv}). Simulations of the strong interactions are by now able to map this intricate behavior of the strong coupling over a wide range of experimentally relevant scales, again leading to excellent agreement between theory and experiment (for a community overview see chap.~9 of \cite{Aoki:2021kgd}).

Going beyond the static or thermodynamic properties of nuclear matter proves to be challenging for both theory and experiment. In heavy-ion collisions most observed particles in the final state at best carry a memory on the whole time-evolution of the collision. This requires phenomenology to disentangle the physics of the QGP from other effects e.g. those arising in the early partonic stages or the hadronic aftermath of the collision. It turns out that in order to construct accurate multi-stage models of the collision dynamics (see e.g. \cite{Lin:2004en,Petersen:2008dd,Bratkovskaya:2011wp}), a variety of first-principles insight is needed. The dynamics of the bulk of the light quarks and gluons which make up the QGP produced in the collision is conveniently characterized by transport coefficients. Of central interest are the viscosities of deconfined quarks and gluons and their color charge conductivity. The physics of hard probes, such as fast jets (see e.g. \cite{Cao:2020wlm}) or slow heavy quark bound states (see e.g. \cite{Rothkopf:2019ipj}), which traverse the bulk nearly as test particles on the other hand requires insight into different types of dynamical quantities. In this context first principles knowledge of the complex in-medium potential between a heavy quark and antiquark, the heavy quark diffusion constant or the so-called jet quenching parameter $\hat q$, which summarizes the momentum broadening of a parton jet is called for. As it turns out computing any of these quantities represents a major challenge for numerical simulation methods of the strong interactions.

Going beyond merely establishing asymptotic freedom and instead revealing the full 6-dimensional phase space (i.e. spatial and momentum distribution) of partons inside nucleons and nuclei is the aim of an ambitious collider project just green-lit in the USA. The upcoming electron-ion collider \cite{AbdulKhalek:2022hcn} will be able to explore the quark and gluon content of nucleons in kinematic regimes previously inaccessible and opens up the first opportunity to carry out precision tomography of nuclei using well-controlled point-particle projectiles. Simulations have already revealed that the virtual particle content of nucleons is vital for the overall angular momentum budget of the proton (see e.g. \cite{Alexandrou:2020sml,Wang:2021vqy}). A computation of the full generalized transverse momentum distribution \cite{Meissner:2009ww} however has not been achieved yet. This quantity describes partons in terms of their longitudinal momentum fraction x, the impact parameter of the collision $b_{\rm T}$ and the transverse momentum of the parton $k_{\rm T}$. Integrating out different parts of the transverse kinematics leads to simpler object, such as transverse momentum distributions (TMDs, integrated over $b_{\rm T}$) or generalized parton distributions (GPDs, integrated over $k_{\rm T}$). Integrating all transverse dependence leads eventually to the conventional parton distribution functions (PDFs), which depend only on the longitudinal Bjorken x variable. A vigorous research community has made significant conceptual and technical progress over the past years, moving towards the first-principles determination of PDFs and more recently GPDs and TMDs from lattice QCD ( for a community overview see \cite{Constantinou:2022yye}). Major advances in the past years include the development of the quasi PDF \cite{Ji:2013dva} and pseudo PDF \cite{Radyushkin:2017cyf} formalism, which offer complementary access to PDFs besides their well-known relation to the hadronic tensor \cite{Liu:1993cv}. With the arrival of the first exascale supercomputer in 2022, major improvements in the precision and accuracy of parton dynamics from lattice QCD are on the horizon.

\subsection{Lattice QCD}

In order to support experiment and phenomenology, theory must provide model independent, i.e. first-principles insight into the dynamics of quarks and gluons in nuclei and within the QGP. This requires the use of quantum chromo dynamics (QCD), the renormalizable quantum field theory underlying the strong interactions. Renormalizability refers to the fact that one only needs to provide a limited number of experimental measurements to calibrate each of its input parameters (strong coupling constant and quark masses) before being able to make predictions at any scale. In order to utilize this vast predictive power of QCD however we must be able to evaluate correlation functions of observables from their defining equations in terms of Feynman's path integral
\begin{align}
\langle O(t_1)\tilde O(t_2)\rangle = \frac{1}{Z} \int {\cal D}[A^\mu_a,\psi^a_f,\bar\psi^a_f] \;O(t_1)\tilde O(t_2)\; {\rm exp}\big[iS_{\rm QCD}[A^\mu_a,\psi^a_f,\bar\psi^a_f]\big],
\end{align}
where $A^\mu_a$ denotes the gluon fields and $\psi^a_f$ the color charged quarks of flavor $f$. The path integral weight is given by the exponentiated QCD action denoted by $S_{\rm QCD}$ (for more details see \cite{Schwartz:2014sze}) and the normalization $Z$ refers to the path integral evaluated in the absence of observables in the integrand.

Computing the dynamical properties of quarks in gluons both inside nucleons as well as in the experimentally accessible QGP requires us to evaluate the above path integral in the presence of strong fluctuations, which invalidate commonly used weak-coupling expansions of the path integral weight. Instead a non-perturbative evaluation of observables is called for. While progress has been made in non-perturbative analytic approaches to QCD, such as the functional renormalization group \cite{Dupuis:2020fhh,Blaizot:2021ikl} or Dyson-Schwinger equations \cite{fischer2006infrared,Roberts:2012sv}, I focus here on the most prominent numerical approach: lattice QCD (for textbooks see e.g. \cite{montvay1994quantum,Gattringer:2010zz}).

In lattice QCD four-dimensional spacetime is discretized on a hypercube with $N^4$ grid points {\bf n}, separated by a lattice spacing $a$. In order to maintain the central defining property of QCD, the invariance of observables under local $SU(3)$ rotations of quark and gluon degrees of freedom, in such a discrete setting, one introduces gauge link variables $U_\mu(x)={\rm exp}[-i g A_\mu^a(x+\frac{1}{2}a\hat\mu)T^a]$, which connect the nodes of the grid in direction $\hat \mu$. Here $g$ denotes the strong coupling constant and $T^a$ refers to the Gell-Mann matrices defining the gauge group $SU(3)$. From the closed products of four or more link variables, as well as the quark fermion fields,  discrete but fully gauge invariant actions can be constructed (the simplest one called the Wilson action). This action allows to formulate a discretized version of Feynman's path integral.

It is the next and final step in the formulation of lattice QCD, which is crucial to understand the challenge we face in extracting dynamical properties from its simulations. The path integral of QCD, while already formulated in a discrete fashion, still contains the canonical complex Feynman weight ${\rm exp}[-iS_{\rm QCD}[U,\psi,\bar\psi]]$. So far, even though progress is being made, no universal numerical method to evaluate such highly dimensional oscillatory integrals has been developed, a challenge often referred to as the sign problem (see e.g. \cite{Gattringer:2016kco,Berger:2019odf}). Instead one circumvents this difficulty by making use of complex analysis and analytically continues the Minkowski time variable $t$ onto the imaginary axis in the lower half complex plane $\tau=it$. The additional factors of the imaginary unit, which arise from this manipulation can be conveniently combined to cancel the pre-factor of $i$ in the Feynman weight leading to
\begin{align}
\langle O_{{\bm n}_1}\tilde O_{{\bm n}_2}\rangle = \frac{1}{Z}\int \prod_{n} \prod_{\mu}dU_{\mu,{\bm n}} d[\psi_{f,{\bm n}},\bar\psi_{f,{\bm n}}] \;O_{{\bm n}_1}\tilde O_{{\bm n}_2}\; {\rm exp}\big[-S_{\rm E}[U,\psi,\bar\psi]\big].
\end{align}
The action $S_{\rm E}\in\mathbb{R}$ one obtains after analytic continuation is referred to as Euclidean action. As a curiosity of quantum field theory one should note that due to a subtle relation between the Boltzmann factor, which describes thermal systems and time evolution in imaginary time, the extent of the imaginary time axis is directly linked to the inverse temperature $\beta=1/T$ of the system (KMS-relation) \cite{Bellac:2011kqa}. By varying the length of the imaginary time axis it is therefore possible to change between a scenario at $T\approx0$ relevant for nucleon structure and $T>0$ relevant for the study of the QGP.

Besides allowing us to incorporate the concept of temperature in a straight forward manner, this Euclidean path integral is now amenable to standard methods of stochastic integration, since the Euclidean Feynman weight is real and bounded from below. Using established Markov-Chain Monte Carlo techniques one generates ensembles of gauge field configurations distributed according to $\frac{1}{Z}{\rm exp}\big[-S_{\rm E}[U,\psi,\bar\psi]\big]$. Evaluating (\textit{measuring}) correlation functions $D(\tau=\tau_2-\tau_1)=\langle O(\tau_1)O(\tau_2)\rangle$ on $N_{\rm conf}$ on statistically independent field realizations $U^{(k)}$ and computing the mean, systematically estimates the quantum statistical expectation value 
\begin{align}
   D(\tau)= \langle O(\tau_1)O(\tau_2) \rangle = \frac{1}{N_{\rm conf}}\sum_{k=1}^{N_{\rm conf}} {O(\tau_1;U^{(k)})O(\tau_2;U^{(k)})} + {\cal O}(1/\sqrt{N_{\rm conf}}).
\end{align}
Here the error decreases with the number of generated configurations independent of the dimensionality of the underlying integral.

To avoid misunderstandings, let me emphasize that results obtained from lattice QCD at finite lattice spacing may not be directly compared to physical measurements. A valid comparison requires that the so-called continuum limit is taken $a\to0$, while remaining close to the thermodynamic limit $V\to\infty$. Different lattice discretizations may yield deviating results, as long as this limit has not been adequately performed. For precision lattice QCD computations a community quality control has been established through the FLAG working group \cite{Aoki:2021kgd} to catalog different simulation results including information on the limits taken.

\section{The Inverse Problem}

The technical challenge we face is now laid bare: in order to make progress in the study of the dynamics of the strong interactions we need to evaluate Minkowski time correlation functions in QCD, related to parton distribution functions in nucleons or the dynamical properties of partons in the QGP. The lattice QCD simulations we are able to carry out however are restricted to imaginary time. Reverting back to the real-time domain as it turns out presents an ill-posed inverse problem.

The key to attacking this challenge is provided by the spectral representation of correlation functions \cite{Bellac:2011kqa}. It tells us that different incarnations of relevant correlation functions (e.g. the retarded or Euclidean correlators) share common information content in the form of a so-called spectral function \cite{Ghiglieri:2020dpq}. The K\"all\'en–Lehmann representation reveals that the retarded correlator of fields in momentum space may be written as
\begin{align}
D^{\rm R}(p_0,{\bf p})=\frac{i}{\pi}\int d\mu \frac{1}{p_0 - \mu+i\epsilon} \rho(\mu,{\bf p}),
\end{align}
while the same correlator in Euclidean time is given as
\begin{align}
D^{\rm E}(\tau,{\bf p})=\int \frac{e^{-\tau \mu}}{1\mp e^{-\beta\mu}} \rho(\mu,{\bf p}).
\end{align}
where the sign in the denominator differs between bosonic $(-)$ and fermionic $(+)$ correlators. Both real-time and Euclidean correlator therefore can be expressed by the same spectral function, integrated over different analytically known kernel functions.

As we do have access to the Euclidean correlator, extracting the spectral function from it in principle gives us direct access to its Minkowski counterpart. It is important to note that often phenomenologically relevant physics is encoded directly and intuitively in the structures of the spectral function, making an evaluation of the real-time correlator superfluous. Transport coefficients e.g. can be read off from the low frequency behavior of the zero-momentum spectral function of an appropriate correlation function \cite{Meyer:2011gj}.

For the extraction of parton distribution function similar challenges ensue. PDFs can be computed from a quantity christened the hadronic tensor $W^{\rm M}(t)$ \cite{Liu:1993cv}, a four-point correlation function of quark fields in Minkowski time. The Euclidean hadronic tensor on the lattice is related to its real-time counterpart via a Laplace transform
\begin{align}
W^{\rm E}(\tau)=\int \,d\mu\, e^{-\mu \tau}\, W^{\rm M}(\mu)
\end{align}
that needs to be inverted. Recently the pseudo PDF approach \cite{Radyushkin:2017cyf} has shown how a less numerically costly three-point correlation function ${\cal M}^{\rm Ioffe}$ can be used to extract similar information on e.g. quark distributions $q(x)$. It too is hidden behind an inverse problem of the form
\begin{align}
{\cal M}^{\rm Ioffe}(\nu)=\int \,dx\,{\rm cos}(\nu x) \,q(x),
\end{align}
where the Ioffe-time matrix elements ${\cal M}^{\rm Ioffe}(\nu)$ are accessible on the lattice.

All the above examples of inverse problems share that in practice they are in fact ill-posed. Not only is the Euclidean correlator from the lattice $D_i$ known only at $N_\tau$ discrete points $\tau_i$, but in addition, as it arises from a Monte-Carlo simulations, it also carries a finite error $\Delta D/D\neq0$. Let us write down the discretized spectral representation in terms of a spectral function $\rho_l$ discretized at frequencies $\mu_l$ along $N_\mu$ equidistant frequency bins of with $\Delta\mu_l$ and the discretized kernel matrix $K_{il}$
\begin{align}
    D^\rho_i = \frac{1}{2}\Delta\mu_1 K_{i1}\; \rho_1 + \sum_{l=2}^{N_\mu-1} \Delta\mu_l K_{il}\; \rho_l +\frac{1}{2}\Delta\mu_{N_\mu} K_{iN_{\mu}}\; \rho_{N_{\mu}}. \label{eq:discrspecrec}
\end{align}
The task at hand is to solve the inverse problem of determining the parameters $\rho_l$ from the sparse and noisy $D_i$'s. The ill-posedness of this inverse problem is manifest in two aspects:

State-of-the-art lattice QCD simulations provide only around ${\cal O}(10-100)$ points along imaginary time $\tau$. From it we must reconstruct the function $\rho$, which often contains intricate patterns at different scales. The fact that $N_\mu\gg N_\tau$ entails that many degenerate sets of $\rho_l$ exist, which all reproduce the input data $D_i$ within their statistical uncertainty. The inverse problem is thus highly degenerate.

In addition many of the kernel functions we have to deal with are of exponential form. This entails a strong loss of information between the spectral function and the Euclidean correlator. In other words, large changes in the spectral function translate into minute changes in the Euclidean correlator. Indeed, each of the tiny eigenvalues of the kernel is associated with a mode along frequencies, which can be added to the spectral function without significantly changing the correlator. Ref.~\cite{Shi:2022yqw} has recently investigated this fact in detail analytically for the bosonic finite temperature kernel relevant in transport coefficient computations.

Even the at first sight benign $cos$ kernel matrix arising in the pseudo PDF approach turns out to feature exponentially diminishing eigenvalues \cite{Karpie:2019eiq} as the lattice simulation cannot access the full Brillouin zone in $\nu$. I.e. the matrix $K_{il}$ is in general ill-conditioned, making its inversion unstable even if no noise is present. In the presence of noise the exponentially small eigenvalues lead to a strong enhancement of even minute uncertainties in the correlation functions rendering the inversion meaningless without further regularization.

We will see in the next section how Bayesian inference can be used to give meaning to the inverse problem arising in extracting real-time dynamics from lattice QCD.

\section{Bayesian Inference of Spectral Functions}

\begin{figure}
    \centering
    \includegraphics[scale=0.5]{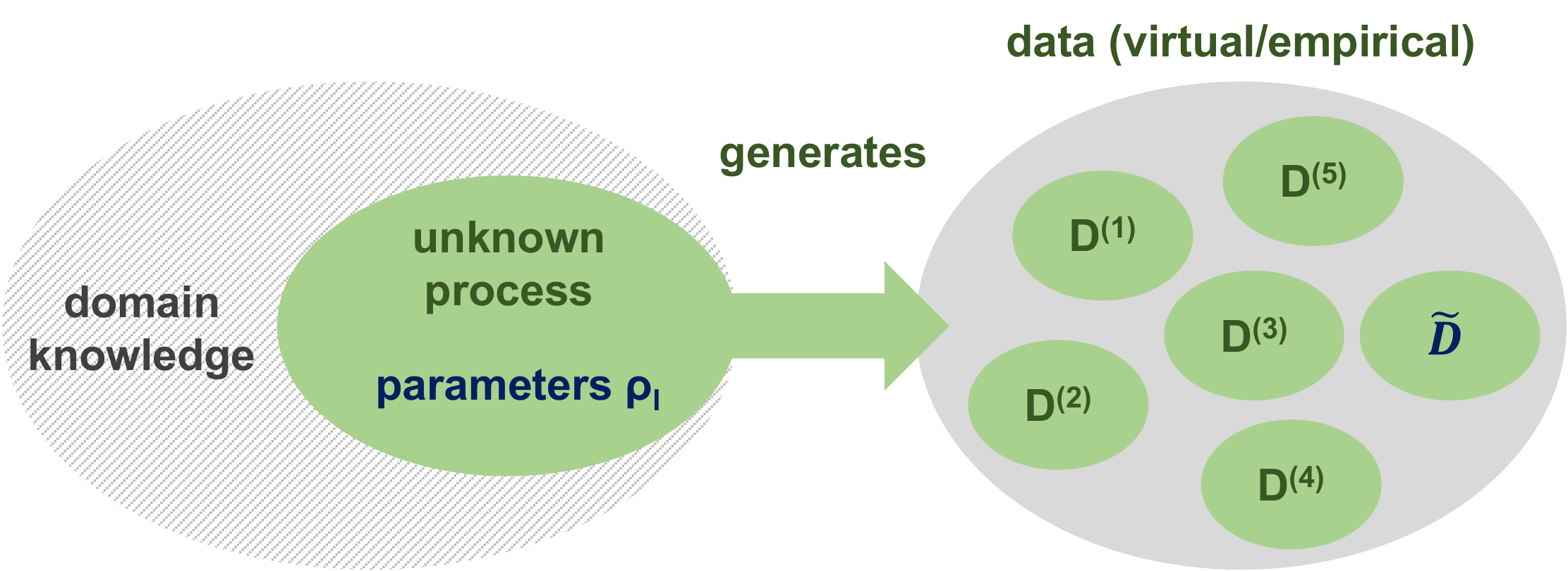}
    \caption{Statistical inference attempts to estimate from observed data $D^{(k)}$ the unknown process parameters $\rho_l$ and as of yet unobserved data $\tilde D$. Bayesian inference exploits the fact that in many instances our model of the unknown process is embedded in a domain from which prior knowledge can be derived.}
    \label{fig:InferenceGen}
\end{figure}

The use of Bayesian inference to extract spectral functions from lattice QCD simulations was pioneered by a team of researchers from Japan in two seminal papers \cite{Nakahara:1999vy,Asakawa:2000tr}. Inspired by prior work in condensed matter physics \cite{jarrell_bayesian_1996} and image reconstruction \cite{skilling1991bayesian}, the team successfully transferred the approach to the extraction of QCD real-time information. The work sparked a wealth of subsequent studies, which have applied and further developed Bayesian techniques to the extraction of real-time information from lattice QCD in various contexts, zero temperature hadron spectra and excited states \cite{PhysRevD.65.014501,SASAKI2005208,PhysRevD.65.094512}, parton-distribution functions \cite{Karpie:2019eiq,Liang:2019frk}, in-medium hadrons \cite{Asakawa:2003re,Datta:2003ww,Umeda:2002vr,Jakovac:2006sf,Aarts:2011sm,Aarts:2012ka,Ding:2012sp,Aarts:2013kaa,Aarts:2014cda,Borsanyi:2014vka,Kim:2014iga,Ikeda:2016czj,Kelly:2018hsi,Kim:2018yhk}, sum rules \cite{Gubler:2011ua,Araki:2014qya}, transport coefficients \cite{Meyer:2007ic,Meyer:2007dy,Aarts:2007wj, Ding:2010ga, Meyer:2011gj,Aarts:2014nba,Amato:2013naa} and the complex in-medium heavy quark potential \cite{Rothkopf:2011db,Burnier:2014ssa,Burnier:2015tda,Burnier:2016mxc}.

The following discussion focuses on the Bayesian extraction of spectral functions that does not rely on a fixed parameterization of the functional form of $\rho$. If strong prior information exists, e.g. if vacuum hadronic spectral functions consist of well separated delta peaks, direct Bayesian parameter fitting methods are applicable \cite{Lepage:2001ym} and may be advantageous. Similarly, some studies of in-medium spectra and transport phenomena deploy explicit parameterizations of the spectral function derived from model input, whose parameters can be fitted in a Bayesian fashion (see Ref.~\cite{Burnier:2017bod} for a recent example). Our goal here is to extract spectral features for systems in which no such apriori parameterization is known.

\subsection{Bayesian inference}

Bayesian inference is a sub-field of statistical data analysis (for an excellent introduction see e.g. \cite{statrethinkingbook,BishopPRML}), which focuses on the estimation of unobserved quantities, based on incomplete and uncertain observed data (see \cref{fig:InferenceGen}). The term unobserved is used to refer to the unknown parameters governing the process, which generates the observed data or to as of yet unobserved future data. In the context of the inverse problem in lattice QCD, the Euclidean correlation functions produced by a Monte-Carlo simulation take on the role of the observed data while the unobserved parameters are the values of the discretized spectral function $\rho_l$. Future observations can be understood as further realizations of the Euclidean correlator along the Markov-Chain of the simulation.

What makes Bayesian inference particularly well suited to attack the inverse problem is that it offers an explicit and well controlled strategy to incorporate information $(I)$ beyond the measured data $(D)$ into the reconstruction of spectral functions $(\rho)$. It does so by using a more flexible concept of probability, which does not necessarily rely on the outcome of a large number of repeatable trials but instead assigns a general degree of uncertainty.

To be more concrete, Bayesian inference asks us to acknowledge that any model of a physical process is constructed within the context of its specific domain, in our case strong interaction physics. I.e. the structure of the model and its parameters are chosen according to prior information obtained within its domain. Bayesian inference then requires us to explicitly assign degrees of uncertainty to all these choices and propagate this uncertainty into a generalized probability distribution called the \textit{posterior} $P[\rho|D,I]$. Intuitively it describes how probable it is that a test function $\rho$ is the correct spectral function, given simulated data $D$ and prior QCD knowledge $I$.

The starting point of any inference task is the joint probability distribution $P[\rho,D,I]$. As it refers to the parameters $\rho$, data $D$ and prior information $I$ it combines information about the specific process generating the data as well as the domain it is embedded in. After applying the rules of conditional probability one obtains the work-horse of Bayesian inference, the eponymous Bayes theorem
\begin{align}
    \underbracket{P[\rho|D,I]}_{\rm posterior} = \underbracket{P[D|I,\rho]}_{\rm likelihood} \underbracket{P[\rho|I]}_{\rm prior} / \underbracket{P[D|I]}_{\rm evidence}.
\end{align}
It tells us how the posterior $P[\rho|D,I]$ can be efficiently computed. The \textit{likelihood} denotes the probability for the data $D$ to be generated from QCD given a fixed spectral function $\rho$. The prior probability quantifies how compatible $\rho$ is compared to our domain knowledge. Historically the $\rho$ independent normalization has been called the \textit{evidence}. Let us construct the different ingredients to Bayes theorem in the following.

What is the likelihood in the case of spectral function reconstruction? Since in Monte-Carlo simulations one usually computes sub-averages of correlation functions on each of the $N_{\rm conf}$ generated gauge field configurations, the data is to a good approximation normal distributed. The corresponding likelihood probability $P[D|\rho,I]\propto {\rm exp}[-L]$, written in terms of the likelihood function $L$, is therefore a multidimensional Gaussian
\begin{align}
    P[D|\rho,I]= {\cal N}[D^\rho,C] \propto {\rm exp}\Big[ - \sum_{ij}\frac{1}{2} (D_i-D^\rho_i)C_{ij}^{-1}(D_j - D^\rho_j)\Big]\label{eq:likelihood},
\end{align}
where $D_i$ denotes the mean of the simulated data at the $i$th Euclidean time step and $D^\rho_i$ the corresponding Euclidean datapoint, arising from inserting the parameters $\rho_l$ into the spectral representation \cref{eq:discrspecrec}. $C_{ij}$ refers to the covariance matrix of the mean
\begin{align}
    C_{ij}=\frac{1}{N_{\rm conf}(N_{\rm conf}-1)} \sum_{k=1}^{N_{\rm conf}} \big( D^{(k)}_i-D_i\big)\big( D^{(k)}_j-D_j\big),
\end{align}
where the individual measurements enter as $D^{(k)}$. Note that in order to obtain an accurate estimate of $C_{ij}$, the number of samples $N_{\rm conf}$ must be significantly larger than the number of data along imaginary time. In particular $C_{ij}$ develops exact zero eigenvalues if the number of configurations is less than that of the datapoints. A speedup in the computation of the likelihood can be achieved in practice if, following Ref.~\cite{Nakahara:1999vy}, one computes the eigenvalues $\sigma_i$ and eigenvectors of $C$ and changes both the kernel and the input data into the coordinate system where $S^t C S= {\rm diag}[\sigma_i]$ becomes diagonal. Then the two sums in \cref{eq:likelihood} collapse onto a single one $L=\sum_{i} \frac{1}{2}(\tilde D_i-\tilde D^\rho_i)/\sigma_i^2$ with $\tilde D^\rho_i = S^t_{ij} K_{jl} \rho_l$ and $\tilde D_i=S^t_{ij}D_j$.

Since the likelihood is a central ingredient in the posterior, all Bayesian reconstruction methods ensure that the reconstructed spectral function, when inserted into the spectral representation will reproduce the input data within their uncertainty. I.e. they will always produce a valid statistical hypothesis for the simulation data. This crucial property distinguishes the Bayesian approach from competing non-Bayesian methods, such as the Backus-Gilbert method and the Pad\'e reconstruction (see examples in \cite{Cyrol:2018xeq}), in which the reconstructed spectral function does not necessarily reproduce the input data.

In case that we do not possess any prior information we have $P[\rho|I]=1$ and Bayes theorem only contains the likelihood. Since the functional $L$ is highly degenerate in terms of $\rho_l$'s, the question of what is the most probable spectral function, i.e. the maximum likelihood estimate of $\rho$, does not make sense at this point. Only by supplying meaningful prior information can we regularize and thus give meaning to the inverse problem.

\subsection{Bayesian spectral function reconstruction}
\label{sec:bayedspecrec}
Different Bayesian strategies to attack the ill-posed spectral function inverse problem differ by the type of domain information they incorporate in the prior probability $P[\rho|I]\propto {\rm exp}[S]$, where $S$ is called the regulator functional. Once the prior probability is constructed, the spectral reconstruction consists of evaluating the posterior probability $P[\rho|D,I]$, which informs us of the distribution of the values of $\rho_l$ in each frequency bin $\mu_l$.

The versatility of the Bayesian approach actually allows us to reinterpret several classic regularization prescriptions in the language of Bayes theorem, providing a unifying language to seemingly different strategies.

When surveying approaches to inverse problems in other fields, \textit{Tikhonov regularization} \cite{tikhonov_stability_1943} is by far the most popular regularization prescription. It amounts to choosing an independent Gaussian prior probability for each parameter
\begin{align}
    P[\rho|I]= \prod_{l=1}^{N_\mu} {\cal N}[m_l,1/\sqrt{\alpha_l}] \propto {\rm exp}\Big[ - \sum_{l=1}^{N_\mu} \alpha_l\; \frac{1}{2}\,(\rho_l-m_l)^2\Big]. \label{eq:tikhonov}
\end{align}
Each normal distribution is characterized by its maximum (mean) denoted here by $m_l$ and width (uncertainty) $1/\alpha_l$. In the literature $m_l$ is usually referred to as the \textit{default model} and $\alpha_l$ simply as hyperparameter. The significance of the two quantities is that in the absence of simulation data, $m_l$ denotes the most probable apriori value of $\rho_l$ with intrinsic uncertainty $1/\alpha_l$. Since these parameters, even though they are constrained by QCD, will be known only up to a some uncertainty, the Bayesian strategy requires us to assign distributions $P[m]$ and $P[\alpha]$ to these model parameters. This is a first example of a so-called hierarchical model, where each level of the model encodes the uncertainties and correlations among model (hyper-)parameters in the subsequent layer. It then remains the task of the user to extract from QCD domain knowledge appropriate uncertainty budgets for $m$ and $\alpha$.

Another regularization deployed in the field of image reconstruction is the so-called \textit{total variation} approach \cite{1992PhyD...60..259R}. Here the difference between neighboring parameters $\rho_l$ and $\rho_{l+1}$ , i.e. $\Delta \rho_l$, is modelled \cite{bardsley2012laplace} as a Laplace distribution 
\begin{align}
    P[\Delta \rho|I]= \prod_{l=1}^{N_\mu-1} {\rm Laplace}[m_l,\alpha_l] \propto {\rm exp}\Big[- \sum_{l=1}^{N_\mu-1}  \alpha_l\; | (\rho_{l+1}-\rho_l) - m_l |^2\Big] \label{eq:tvreg}.
\end{align}
Since $\Delta \rho_l$ is related to the first derivative of the spectral function this regulator incorporates knowledge about rapid changes, such as kinks, in spectral features. Choosing $\alpha_l$ and $m_l$ appropriately one may e.g. prevent the occurrence of kink features in the reconstructed spectral function, if it is known that the underlying true QCD spectral function is smooth.

In Ref.~\cite{Fischer:2017kbq} I proposed a regulator related to the derivative of $\rho$, with a different physical meaning
\begin{align}
    P[\Delta \rho|I]= \prod_{l=1}^{N_\mu-1} {\cal N}[m_l,\alpha_l] \propto {\rm exp}\Big[- \sum_{l=1}^{N_\mu-1}  \alpha_l\; \Big( (\rho_{l+1}-\rho_l) - m_l \Big)^2\Big].
\end{align}
Often spectral reconstructions, which are based on a relatively small number of input data, suffer from \textit{ringing artifacts}, similar to the Gibbs ringing arising in the inverse problem of the Fourier series. These artifacts lead to a reconstructed spectral function with a similar area as the true spectral function but with a much larger arc length due to the presence of unphysical wiggles. Since such ringing is not present in the true QCD spectral function we may apriori suppress it by penalizing arc length $\ell=\int d\mu\; \sqrt{1+(d\rho/d\mu)^2}$. And since the square root is monotonous, we may remove it for our purposes, as well as, discard the addition of unity, as it is absorbed into the normalization of the corresponding prior distribution. The hyperparameters of such a prior must be chosen appropriately, since the remedy to one artifact, ringing, can lead to the introduction of a different artifact, which is \textit{over-damping} of reconstructed spectral features. The relevant ranges for $\alpha$ and $m$, as e.g. in Ref.~\cite{Kim:2018yhk}, can be established using mock data tests.

If our prior domain knowledge contains information about the smoothness and the absence of ringing then it is of course possible to combine different regulators by multiplying the prior probabilities. The reconstruction of the first picture of a black hole e.g. combined the Tikhonov and total variation regularization \cite{2019ApJ...875L...1E}. In the presence of multiple regulators, the hyperparameters $\alpha$ and $m$ of each of these distributions need to be assigned an (independent) uncertainty distribution. 

One may ask, whether a proliferation of such parameters spoils the benefit of the Bayesian approach? The answer is that in practice one can estimate the probable ranges of these parameters by use of mock data. One carries out the spectral function reconstruction, i.e. the estimation of the posterior probability $P[\rho|D,I]$, using data, which has been constructed from known spectral functions with realistic features and which has been distorted with noise similar to those occurring in Monte-Carlo simulations (see e.g. \cite{Kim:2018yhk}). One may then observe from such test data sets, what the most probable values of the hyperparameters are and in what interval they vary, depending on different spectral features present in the input data.

The three priors discussed so far are not commonly used as stand-alone regulators in the reconstruction of QCD spectral functions in practice. The reason is that neither of them can exploit a central prior information available in the lattice context, which is the positivity \cite{montvay1994quantum} of the most relevant hadronic spectral functions\footnote{For the reconstruction of non-positive spectral functions using Bayesian approaches see e.g. the Hobson-Lasenby modification of the MEM \cite{Hobson:1998bz}, the shift strategy of \cite{Haas:2013hpa} or my extension of the BR method \cite{Rothkopf:2016luz}.}. I.e. in most of the relevant reconstruction tasks from lattice QCD, the problem can be formulated in terms of a positive definite spectral function, which significantly limits the function space of potential solutions. Methods that are unable to exploit this prior information, such as the Backus-Gilbert method have therefore been shown to perform poorly relative to the Bayesian approaches, when it comes to the reconstruction of well-defined spectral features (see e.g. \cite{Liang:2019frk}).

In the following let us focus on two prominent Bayesian methods, which have been deployed in the reconstruction of positive spectral functions from lattice QCD, the \textit{Maximum Entropy Method} (MEM) and the \textit{Bayesian Reconstruction} (BR) method. 

The MEM \cite{narayan1986maximum,skilling1991bayesian,jarrell_bayesian_1996,Asakawa:2000tr} has originally been constructed to attack image reconstruction problems in astronomy. It therefore focuses on two-dimensional input data and deploys the Shannon-Jaynes entropy $S_{\rm SJ}$ as regulator:
\begin{align}
    P[\rho|I]\propto {\rm exp}\Big[ - \sum_{l=1}^{N_\mu}  \alpha_l\; \Delta\mu \Big(\rho_l-m_l-\rho_l {\rm log}\Big[ \frac{\rho_l}{m_l} \Big] \Big)\Big]. 
\end{align}
Its regulator is based on four axioms \cite{skilling1991bayesian}, which specify the prior information the method exploits. They are \textit{subset independence}, which states that prior information on $\rho_l$'s at different discrete frequency bins $l$ can be combined in a linear fashion within $S_{\rm SJ}$. The second axiom enforces that $S_{\rm SJ}$ has its \textit{maximum at the default model}, which establishes the meaning of $m_l$ as the apriori most probable value of $\rho_l$ in the absence of data. These two axioms are not specific to the MEM and find use in different Bayesian methods. It is the third and fourth axiom that distinguish the MEM from other approaches: \textit{coordinate invariance} requires that $\rho$ itself should transform as a dimensionless probability distribution and \textit{system independence} assumes certain factorizability properties of a two-dimensional spectral function along the two dimensions. 

From the appearance of the logarithm in $S_{\rm SJ}$ it is clear that the MEM can exploit the positivity of the spectral function. Due to the fact that the logarithm is multiplied by $\rho$, $S_{\rm SJ}$ is actually able to accommodate exact zero values of a spectral function. Since the reconstruction task in lattice QCD is one-dimensional, it is not obvious how to directly translate system independence. An intuitive way of interpreting this axiom using e.g. the monkeys and kangaroos example of Ref.~\cite{jarrell_bayesian_1996} is that the MEM shall not introduce correlations among $\rho_l$'s where the data does not require it. This is a quite restrictive property, as it is exactly prior information, which should help us to limit the potential solutions space by providing as much information about the structure of $\rho$ as possible. Similarly, the assumption that $\rho$ must transform as a probability distribution, while appropriate for a distribution of dimensionless pixel values in an image, does not necessarily apply to spectral functions. These are in general dimensionful quantities and may even contain UV divergences when evaluated naively.

To overcome these conceptual difficulties the \textit{BR method} was developed in Ref.~\cite{Burnier:2013nla} with the one-dimensional reconstruction problem of lattice QCD real-time dynamics in mind. The BR method features a regulator $S_{\rm BR}$ related to the Gamma distribution 
\begin{align}
    \nonumber P[\rho|I]=&\;\prod_{l=1}^{N_\mu}{\rm Gamma}[1+\Delta\mu\alpha_l,\Delta\mu\alpha_l/m_l], \\
    \propto&\; {\rm exp}\Big[ - \sum_{l=1}^{N_\mu}  \alpha_l\;\Delta\mu\; \Big(1-\frac{\rho_l}{m_l}-{\rm log}\Big[ \frac{\rho_l}{m_l} \Big] \Big)\Big],
\end{align}
which looks similar to the Shannon-Jaynes entropy but differs in crucial ways. Its construction shares the first two axioms of the MEM but replaces the third and fourth axiom with the following: \textit{scale invariance} enforces that the posterior may not depend on the units of the spectral function, leading to only ratios between $\rho_l$ and the default model $m_l$, which by definition must share the same units. The use of ratios also requires that neither $\rho$ nor $m$ vanishes. $S_{\rm BR}$ differs therefore from the Shannon-Jaynes regulator where the integrand of $S_{\rm SJ}$ is dimensionful. The units of $\Delta_\mu$ enter as multiplicative scale and can be absorbed into a redefinition of $\alpha$ (and which will be marginalized over as described in \cref{sec:uncertmap}). Furthermore, one introduces a $\textit{smoothness}$ axiom, which requires the spectral function to be twice differentiable. While it may appear that the latter axiom is at odds with the potential presence of delta-function like structures in spectral functions, it ensures that one smoothly approximates such well defined peaks as the input data improves.

\begin{figure}
    \centering
    \includegraphics[scale=0.4]{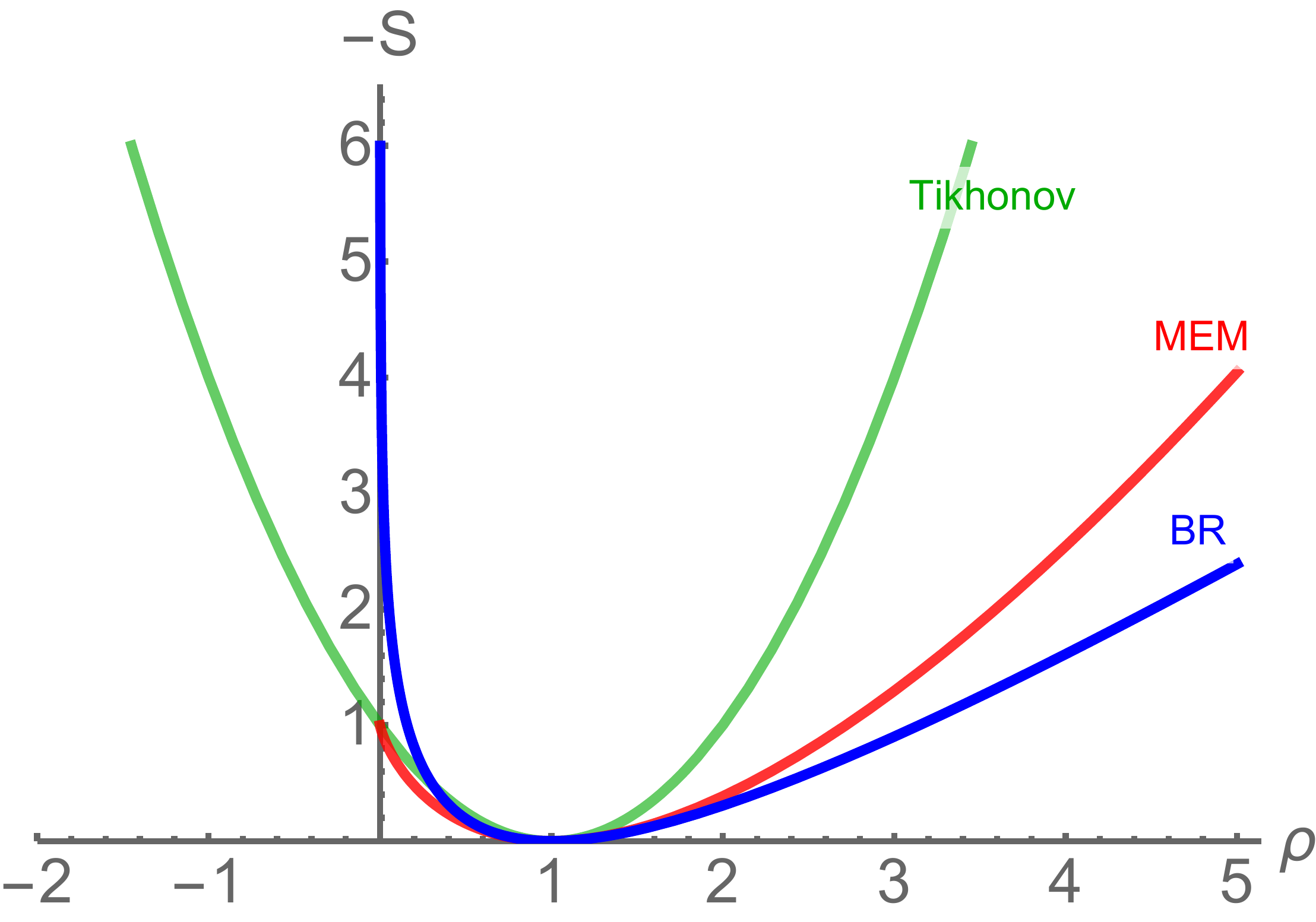}
    \includegraphics[scale=0.4]{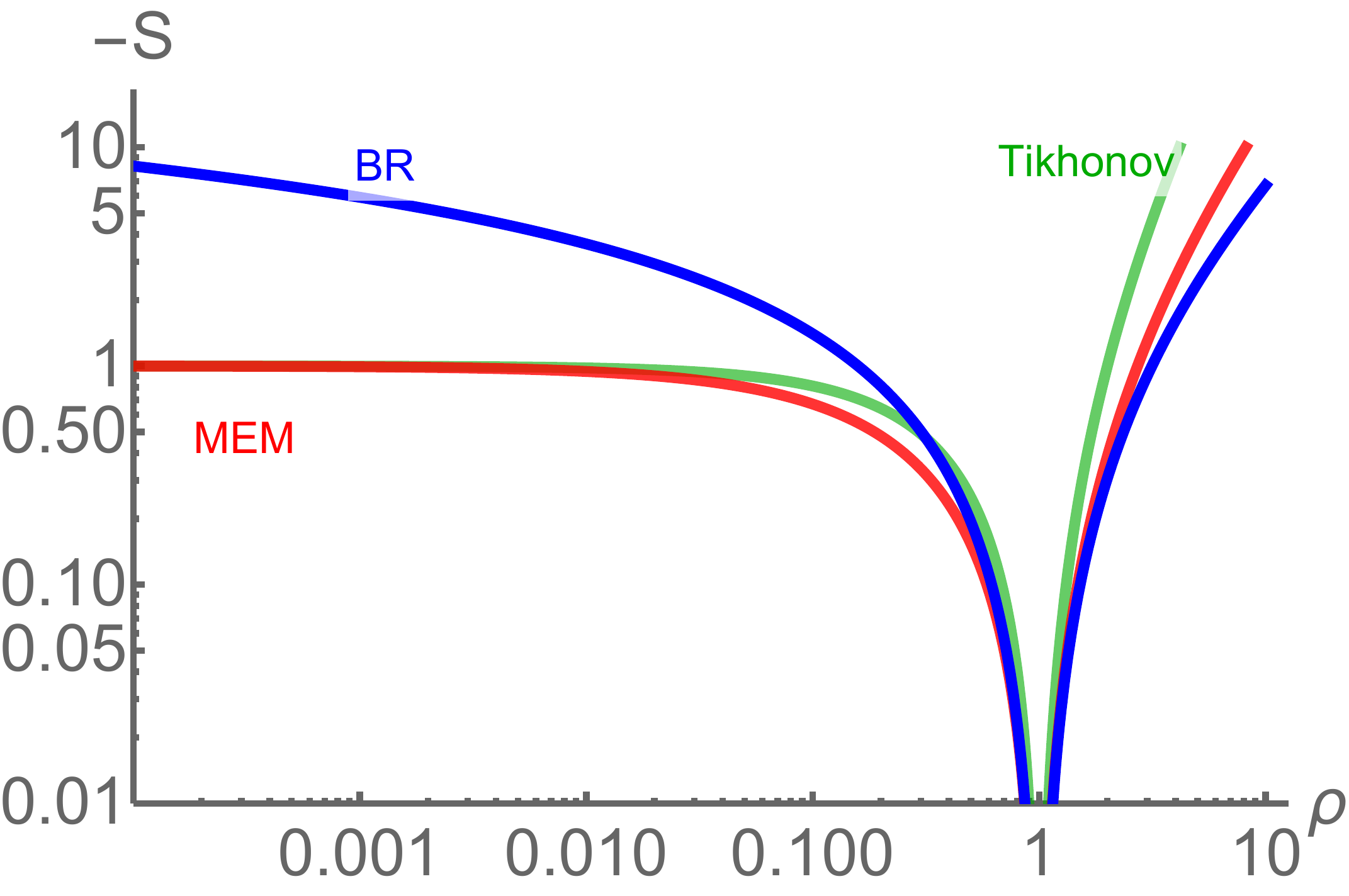}
    \caption{Comparison of the regulators of the Tikhonov approach (green), the MEM (red) and the BR method (blue) in linear scale (top) and double logarithmic scale (bottom). The Shannon-Jaynes regulator accommodates $\rho=0$ but appears flat for spectral functions with values close to zero. The BR prior shows the weakest curvature for $\rho>m$ among all regulators.}
    \label{fig:regulators}
\end{figure}

Let us compare the regulators of the Tikhonov approach, the MEM and the BR method in \cref{fig:regulators}, which plots the negative of the integrand for the choice of $m=1$. The top panel shows a linear plot, the bottom panel a double logarithmic plot. By construction, all feature an extremum at $\rho=m$ and the MEM and BR enforce positive (semi-)definiteness of the spectral function. The functional form of the BR regulator turns out to be the one with the weakest curvature among all three for $\rho>m$, while it still manages to regularize the inverse problem. Note that the weaker the regulator, the more efficiently it allows information in the data to manifest itself (it is actually the weakest on the market). At the same time a weaker regulator is less potent in suppressing artifacts, such as ringing, which may affect spectral function reconstruction based on very small number of datapoints\footnote{To avoid this complication, the BR regulator has been successfully combined with the arc-length penalty regulator in Ref.~\cite{Kim:2018yhk}.}. 

At this point we are ready to carry out the Bayesian spectral reconstruction. I.e. after choosing according to one's domain knowledge a prior distribution $P[\rho|I(m,\alpha)]$ and assigning appropriate uncertainty intervals to their hyperparameters $P[\alpha]$ and $P[m]$ via mock-data studies, we can proceed to evaluate the posterior distribution $P[\rho|D,I]$. If we can access this highly dimensional object through a Monte-Carlo simulation (see e.g. \cref{sec:BRMC}) it provides us not only with the information of what the most probable spectral function is, given our simulation data, but also contains the complete uncertainty budget, including both statistical (data related) and systematic errors (hyperparameter related). The maximum of the prior defines the most probable value for each $\rho_l$ and its spread allows a robust uncertainty quantification beyond a simple Gaussian approximation (i.e. standard deviation) as it may contain tails that lead to a deviation of the mean from the most probable value.

\subsection{Uncertainty quantification for point estimates}
\label{sec:uncertmap}
While access to the posterior allows for a comprehensive uncertainty analysis, a full evaluation of $P[\rho|D,I]$ historically remained computationally prohibitive. Thus the community focused predominantly\footnote{A few works have explored stochastic strategies for the evaluation of the posterior in the context of the SOM \cite{mishchenko2000diagrammatic} or the stochastic analytic continuation (SAI) method \cite{ding2018stochastic,Shao:2022yez}, of which the MEM is a special limit \cite{beach2004identifying}.} on determining a point estimate of the most probable spectral function from the posterior $P[\rho|D,I]$, also called MAP, the maximum aposteriori estimate
\begin{align}
   \left. \frac{\delta }{\delta \rho}P[\rho|D,I]\right|_{\rho=\rho^{\rm MAP}}=0\label{eq:BayesOpt}.
\end{align}
In this case, while much easier to handle as a numerical optimization task, only a fraction of the information contained in the posterior is made accessible. In particular most information related to uncertainty remains unknown and thus needs to be approximated separately.

The above optimization problem in general can be very demanding as the posterior may contain local extrema in addition to the global one that defines $\rho^{\rm MAP}$. At least in the case of the Tikhonov, MEM and BR methods however it is possible to prove that if an extremum for \cref{eq:BayesOpt} exists it must be unique. The reason is that all three regulators are convex. The proof of this statement does not rely on a specific parameterization of the spectral function and therefore promises that standard (quasi)Newton methods, such as Levenberg-Marquardt or LBFGS (see e.g. Ref.~\cite{10.5555/1403886}) can be used to locate this unique global extremum in the $N_\mu$ dimensional search space.

Also from an information point of view it is fathomable that at this point a unique solution to the former ill-posed inverse problem can be found. We need to estimate the most probable values of $N_\mu$ parameters $\rho_l$ and have now provided $N_\tau$ simulation data $D_i$, as well as $N_\mu$ pieces of information in the form of the $m_l$'s and $\alpha_l$'s each. I.e. the number of knowns $2 N_\mu+N_\tau > N_\mu$ is larger than the number of unknowns, making a unique determination possible. The proof presented in Ref.~\cite{Asakawa:2000tr} formalizes this intuitive statement.

In practice it turns out that the finite intercept of the Shannon-Jaynes entropy for $\rho=0$ can lead to slow convergence if spectral functions with wide ranges of values close to zero are reconstructed. In lattice QCD this occurs regularly when e.g. hadronic spectral functions contain sharp and well separated peak structures. $S_{\rm SJ}$ for very small values (see \cref{fig:regulators}) is effectively flat and thus unable to efficiently guide the optimizer toward the unique minimum and convergence slows down. It is therefore that one finds in the literature that the extremum \cref{eq:BayesOpt} in the MEM is accepted for tolerances around $\Delta\approx 10^{-7}$, which is much larger than zero in machine (double-)precision. Such a large tolerance does not guarantee bitwise identical results when starting the optimization from different initial conditions. The BR prior on the other hand does not exhibit a finite intercept at $\rho=0$ and therefore avoids this slow convergence problem. It has been found to be capable of locating the unique extremum $\rho^{\rm MAP}$ in real-world settings down to machine precision, which guarantees that the reconstruction result is independent of the starting point of the optimizer.

Bayesian inference forces us to acknowledge two sources of uncertainty: \textit{statistical uncertainty} in the data and uncertainty associated to the \textit{choice and parameters of the prior probability}. 

Before continuing to the technical details of how to estimate uncertainty, let us focus on the role of prior information first. It enters both through the selection of a prior probability and the choice of the distributions $P[m]$ and $P[\alpha]$. It is important to recognize that already from an information theory viewpoint, one needs to supply prior information if the goal is to give meaning to an ill-posed inverse problem: originally we started out to estimate $N_\mu\gg N_\tau$ parameters $\rho_l$ from $N_\tau$ noisy input data $D_i$.

I.e. in order to select among the infinitely many degenerate parameter sets $\rho_l$ a single one as the most probable, we need information beyond the likelihood. Conversely any method that offers a unique answer to the inverse problem utilizes some form of prior information, whether it acknowledges it or not. Bayesian inference, by making the role of prior knowledge explicit in Bayes theorem, allows us to straight forwardly explore the dependence of the result on our choices related to domain information. It is therefore ideally suited to assess the influence of prior knowledge on reconstructed spectral functions. This distinguishes it from other approaches, such as the Backus Gilbert method, where a similarly clear distinction of likelihood and prior is absent. The Tikhonov method is another example. Originally formulated with a vanishing default model, one can find statements in the literature that it is default model independent. Reformulated in the Bayesian language, we however understand that its original formulation just referred to one specific choice of model, which made the presence of prior knowledge hard to spot.

\begin{figure}
    \centering
    \includegraphics[scale=0.3]{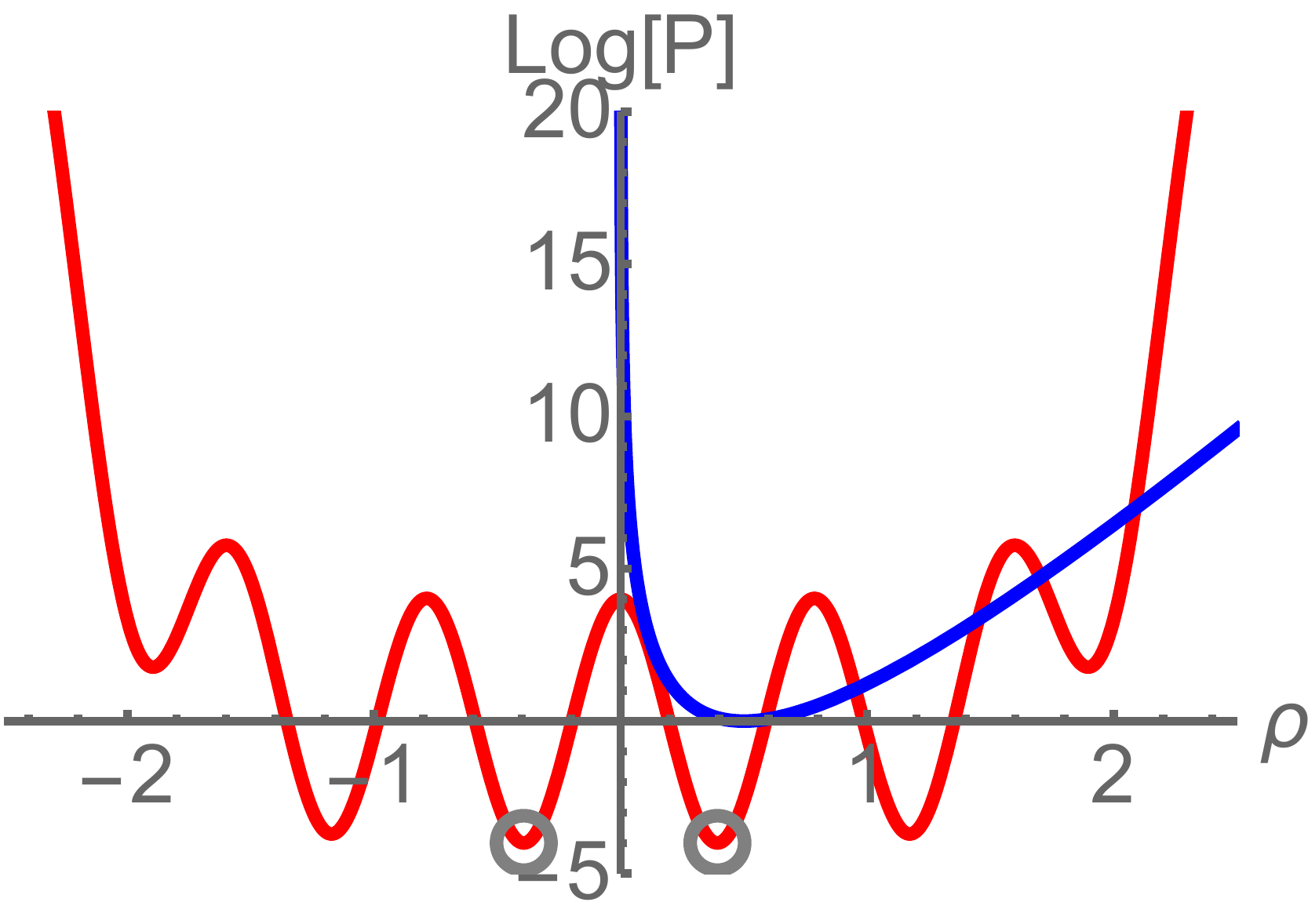}
    \includegraphics[scale=0.3]{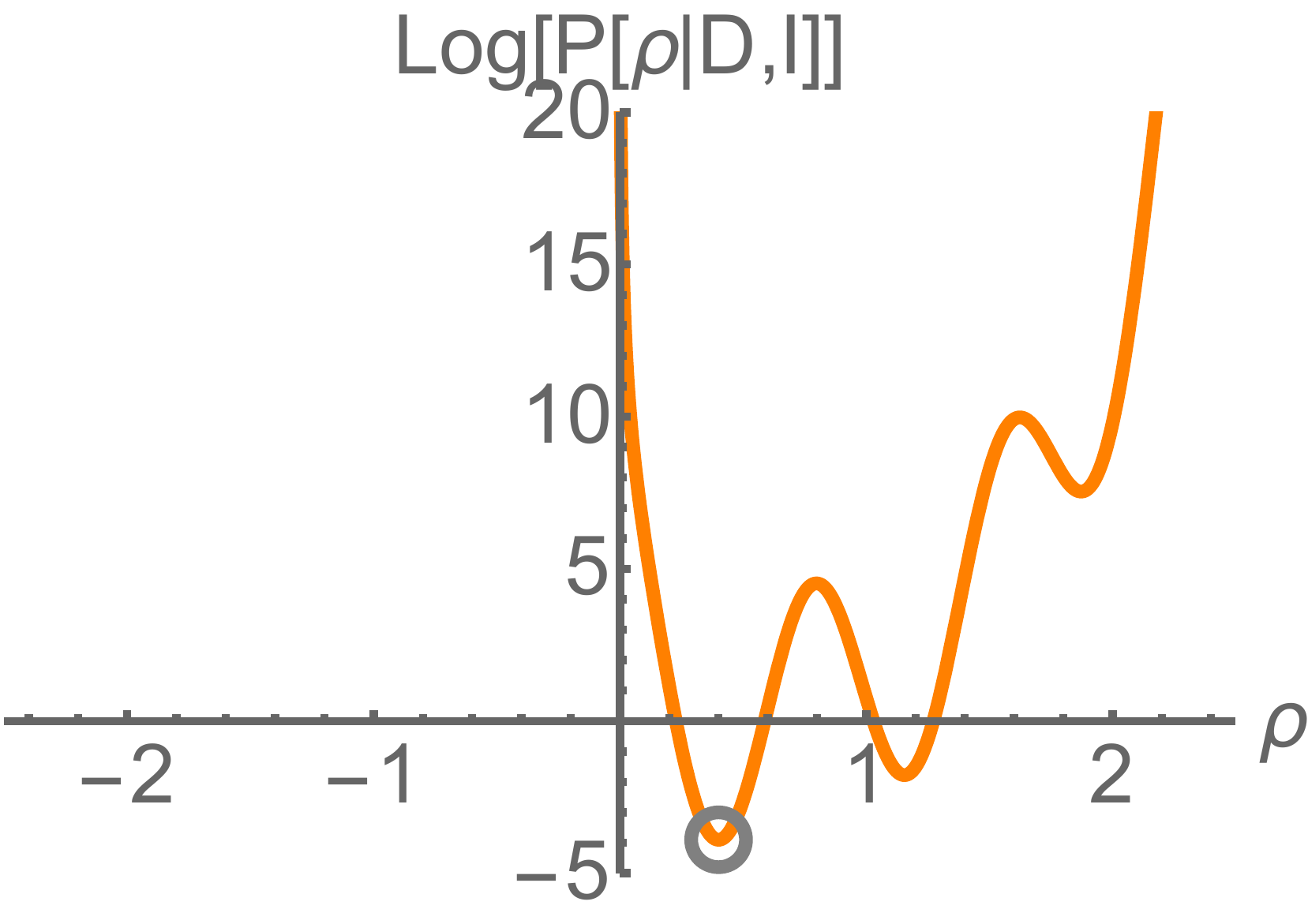}
    \caption{Sketch of how the confluence of (left) likelihood (red) and (a convex) prior (blue) in the posterior (orange, right) leads to a regularization of the inverse problem. Instead of multiple degenerate minima in the likelihood (gray circles) only a single unique one remains in the posterior.}
    \label{fig:sketch1}
\end{figure}

The presence of the prior as regulator also entails that among the structures in a reconstructed spectral function only some are constrained by the simulation data and others are solely constrained by prior information. It is only in the \textit{Bayesian continuum limit}, which refers to taking simultaneously the error on the input data to zero while increasing the number of available datapoints toward infinity, that the whole of the spectral function is fixed by input data alone. Our choice of regulator determines how efficiently we converge to this limit and which type of artefacts (e.g. ringing or over-damping) one will encounter on the way. One important element of uncertainty analysis in Bayesian spectral reconstruction therefore amounts to exploring how reconstructed spectra improve as the data improves\footnote{In lattice QCD it is often easier to collect more samples than to simulate on grids with more points along Euclidean time. Then at least the improvement of the reconstruction with increasing statistics needs to be considered.}. This is a well-established practice in the community.

When reconstructing the spectral function according to a given set of Monte-Carlo estimates $D^{k}_i$ of a lattice QCD correlator $D_i$, we need to reliably estimate the statistical and systematic uncertainty budget. It is important to recognize that these may be related, e.g. increasing the precision of input data often makes the reconstructed spectrum less susceptible to changes in $m$ or $\alpha$. An often deployed strategy is to nevertheless estimate the effects separately: In order to assess statistical uncertainty we may use established bootstrap methods or the (blocked) Jackknife (for an introduction see Ref.~\cite{EfroTibs93}), where the reconstruction is performed repeatedly on subensembles of the input data $D^{k}_i$ and the variance among the reconstructed spectra provides a direct estimate of their statistical uncertainty. 

In the case of point estimates, one usually decides apriori on a regulator and fixes to a certain value of the default model $m$ and of the hyperparameter $\alpha$, before carrying out the reconstruction. The freedom in all these choices enters the systematic uncertainty budget. 

Often the user has access to a reliable default model $m(\omega)$ only along a limited range of frequencies $\mu$. In lattice QCD such information is often obtained from perturbative computations describing the large frequency and momentum behavior of the spectral function. In the low frequency part of the spectrum, where non-perturbative physics dominates, we often do not possess any relevant information about the functional form of $\rho$. It is then customary to extend the default model into the non-perturbative regime using simple and smooth functional forms that join up in the perturbative regime.

In practice the user repeats the reconstruction using different choices for the unknown parts of $m$, e.g. different polynomial dependencies on the frequency and subsequently uses the variation in the end result as indicator of the systematic uncertainty. It is important to note that if there exist different regulators that encode compatible and complementary prior information that one should also consider repeating the reconstruction based on different choices of $P[\rho|I]$ itself. 

Since we have access to the likelihood and prior, we may ask whether a combined estimation of the statistical and systematic uncertainty can be carried out even in the case of a point estimate. Since the reconstructed spectrum $\rho^{\rm MAP}$ denotes a minimum of the posterior, one may try to compute the curvature of the (log) posterior $L-S$ around that minimum, which would indicate how steep or shallow that minimum actually is. This is the strategy laid out e.g. in Ref.~\cite{Asakawa:2000tr}. In practice it relies on a saddle point approximation of the posterior and therefore can lead to an underestimation of the uncertainty. Many recent studies thus deploy a combination of the Jackknife and a manual variation of the default model.

Since the treatment of hyperparameters differs among the various Bayesian methods, let me discuss it here in more detail. Appropriate ranges for the values of $m$ can often be estimated from mock data studies and since the functional dependence of the default model is varied as part of the uncertainty estimation discussed above, we focus here on the treatment of $\alpha$. I.e. we will treat the values of $m$ as fixed and consider the effect of $P[\alpha]$. If alpha is taken to be small, a large uncertainty in the value $m$ ensues, which leads to a weak regularization and therefore to large uncertainty in the posterior. If $\alpha$ is large it constrains the posterior to be close to the prior and limits the information that data can provide to the posterior.

Three popular strategies are found in the literature to treat $\alpha$. Note that in the context of the MEM, a common value is assigned to all hyperparameters $\alpha_l$, i.e. the same uncertainty is assigned to the default model parameters $m_l$ at all frequencies, an ad-hoc choice.

The simplest treatment of $\alpha$, also referred to as the \textit{Morozov criterion} or \textit{historic MEM} is motivated by the goal to avoid over fitting of the input data. It argues that if we knew the correct spectral function and were to compute the corresponding likelihood function $L$, it would on average evaluate to $\langle L\rangle = \frac{1}{2} N_\tau$ i.e. half the number of datapoints. Therefore one should tune the value of $\alpha$ such that the likelihood reproduces this value.

The second and third strategy are based directly on Bayes theorem. The Bayesian way of handling uncertainties in model parameters is to make their dependence explicit in the joint probability distribution $P[\rho,D,I(m,\alpha)]$. Now that the distribution depends on more than three elements, application of conditional probabilities leads to
\begin{align}
    \nonumber P[\rho,D,\alpha,m] =& P[D|\rho,\alpha,m]P[\rho|\alpha,m]P[\alpha,m],\\
    =&P[\alpha|\rho,D,m] P[\rho|D,m]P[D,m]\label{eq:BayesAllExpl}.
\end{align} 

The \textit{modern MEM} approach solves \cref{eq:BayesAllExpl} for $P[\alpha|\rho,D,m]$. It then integrates point estimates $\rho^{\rm MAP}_\alpha$ obtained for fixed values of $\alpha$ over that probability distribution. In order to compute $P[\alpha|\rho,D,m]$ two ingredients are necessary: the full posterior $P[\rho|D,\alpha,m]$ and the distribution $P[\alpha]$. The former is in general not analytically known and therefore is in practice approximated by a saddle point approximation. The latter is in the literature either chosen as constant or as $P[\alpha]\propto 1/\alpha$, a choice referred to as Jeffrey's prior.

Let me briefly clarify the often opaque notion of Jeffrey's prior \cite{doi:10.1098/rspa.1946.0056}. Given a probability distribution $P[x|{\bm \alpha},{\bf m}]$ and a choice of parameter, e.g. $\bm \alpha$, Jeffrey's prior refers to the unique distribution $P_{\rm J}[{\bm \alpha}]=\sqrt{{\rm det}[I({\bm \alpha})]}$ defined from the Fisher information matrix $I({\bm \alpha})$. This definition is considered to be uninformative, as it remains invariant under a change of coordinates of $\bm \alpha$. Using the one-dimensional Gaussian distribution as example, we can obtain an intuitive understanding of its role. Let $P[x|\sigma,m]={\cal N}[x|\sigma,m]$, then 
\begin{align}
P_{\rm J}[m]&=\sqrt{\int dx {\cal N}[x|\sigma,m] \big( \frac{d}{d m} {\cal N}[x|\sigma,m]  \big)^2}=\sqrt{\frac{1}{\sigma^2}}={\rm const.},\\
P_{\rm J}[\sigma]&=\sqrt{\int dx {\cal N}[x|\sigma,m] \big( \frac{d}{d \sigma} {\cal N}[x|\sigma,m]  \big)^2}=\sqrt{\frac{2}{\sigma^2}}=\sqrt{2}\frac{1}{\sigma}.
\end{align}
Jeffrey's prior for $m$ is independent of $m$ and thus refer to the unique \textit{translation invariant} distribution on the real values (Haar-measure for addition). It therefore does not impart any information on the location of the peak of the normal distribution. Similarly $P_{\rm J}[\sigma]$ is a \textit{scale invariant} distribution on the positive real values (Haar-measure for multiplication). Since the uncertainty parameter $\sigma$ enters as a multiplicative scale in the normal distribution its Jeffrey's prior also does not introduce any additional information. Both priors investigated here are improper distributions, i.e. they are well-defined only in products with proper probability distributions. 

The third strategy to treat the parameters $\alpha_l$ has been put forward in the context of the \textit{BR method}. It sets out to overcome the two main limitations of the MEM approach: the need for saddle point approximations in the handling of $\alpha$ and the overly restrictive treatment of assigning a common uncertainty to all $m_l$'s. The BR method succeeds in doing so, by using Bayes theorem to marginalize the parameters $\alpha_l$ apriori, making the (highly conservative) assumption that no information about $\alpha_l$ is known, i.e. $P[\alpha_l]=1$. It benefits from the fact that in contrast to the Shannon-Jaynes prior, the BR-prior is analytically tractable and its normalization can be expressed in closed form.

We start from \cref{eq:BayesAllExpl} and assume that the parameters $\alpha$ and $m$ are independent, so that their distributions factorize. Marginalizing a parameter simply means integrating the posterior over the probability distribution of that parameter. Via application of conditional probabilities it is possible to arrive at the corresponding expression
\begin{align}
    \nonumber \prod_l\int d\alpha_l P[\alpha|\rho,D,m] P[\rho|D,m] &= \frac{P[D|\rho,I]}{P[D|m]P[m]}\prod_l\int d\alpha_l P[\rho|\alpha,m]P[\alpha]P[m],\\
    P[\rho|D,m]&=\frac{P[D|\rho,I]}{P[D|m]}\prod_l\int d\alpha_l P[\rho|\alpha,m]P[\alpha],
\end{align}
where $P[\rho|D,m]$ does not depend on $\alpha$ anymore and by definition of probabilities $\int d\alpha P[\alpha|\rho,D,m]=1$. The posterior $P[\rho|D,m]$ now includes all effects arising from the uncertainty of $\alpha$ without referring to that variable anymore. Due to the form of the BR prior $P[\rho|\alpha,m]$, the integral over $\alpha_l$ is well defined, even though we used the improper distribution $P[\alpha]=1$. One may wonder whether integrating over $\alpha_l$ impacts the convexity of the prior. While not proven rigorously, in practice it turns out that the optimization of the marginalized posterior $P[\rho|D,m]$ in the BR method does not suffer from local extrema.

A user of the BR method therefore only needs to provide a set of values for the default model $m_l$ to compute the most probable spectral function \begin{align}
   \left. \frac{\delta }{\delta \rho}P[\rho|D,m]\right|_{\rho=\rho^{\rm MAP}_{\rm BR}}=0\label{eq:BROpt}.
\end{align}
By carrying out several reconstructions by varying the functional form of $m$ within reasonable bounds, established by mock-data tests, the residual dependence on the default model can be quantified.

So far we have discussed the inherent uncertainties from the use of Bayesian inference and how to assess them. Another source of uncertainty in spectral reconstructions arises from specific implementation choices. Let me give an example based on the Maximum Entropy Method. In order to save computational cost, the MEM historically is combined with a singular value decomposition to limit the dimensionality of the solution space. The argument by Bryan \cite{bryan_maximum_1990} suggests that instead of having to locate the unique extremum of $P[\rho|D,I]$ in the full $N_\mu$ dimensional search space of parameters $\rho_l$, it is sufficient to use a certain parameterization of $\rho(\omega)$ in terms of $N_\tau$ parameters, the number of input data points. The basis functions are obtained from a singular value decomposition (SVD) of the transpose of the kernel matrix $K^t$. Bryan's argument only refers to the functional form of the Kernel $K$ and the number of data points $N_\tau$ in specifying the parameterization of $\rho(\omega)$. If true in general, this would lead to an enormous reduction in computational complexity. However, I have put forward a counter example to Bryan's argument (originally in \cite{Rothkopf:2011ef}) including numerical evidence, which show that in general the extremum of the prior is not part of Bryan's reduced search space. 

One manifestation of the artificial limitation of Bryan's search space is a dependence of the MEM resolution on the position of a spectral feature along the frequency axis. As shown in Fig.3 of Ref.~\cite{Rothkopf:2012vv}, if one reconstructs a single delta peak located at different positions $\mu_0$ with the MEM, one finds that the reconstructed spectral functions show a different width, depending on the value of $\mu_0$. This can be understood by inspecting the SVD basis functions, which are highly oscillatory close to $\mu_{\rm min}$ the smallest frequency chosen to discretize the $\mu$ range. At larger values of $\mu$ these functions however damp towards zero. I.e. if the relevant spectral feature is located in the $\mu$ range where the basis functions have structure, it is possible to reconstruct a sharp peak reasonable well, while if it is located at larger $\mu$ the resolution of the MEM decreases rapidly. The true Bayesian $\rho^{\rm MAP}$, i.e. the global extremum of the MEM posterior, however does not exhibit such a resolution restriction, as one can see when changing the parameterization of the spectral function to a different basis, e.g. the Fourier basis consisting of cos and sin functions. In addition Ref.~\cite{Jakovac:2006sf} in its Fig.28 showed that using a different parameterization of the spectral function, which restricts $\rho$ to a space that is equivalent to the SVD subspace from a linear algebra point of view, one obtains a different result. This, too, emphasizes that the unique global extremum of the posterior is not accessible within these restricted search spaces. Note that one possible explanation for the occurrence of the extremum of $P[\rho|D,I]$ outside of the SVD space lies the fact that in constrained optimization problems (here the constraint is positivity), the extremum can either be given by the stationarity condition of the optimization functional in the interior of the search space or it can lie on the boundary of the search space restricted by the constraint.

I.e. in addition to artifacts introduced into the reconstructed spectrum via a particular choice of prior distribution and handling of its hyperparameters (e.g. ringing or over-damping), one also must be aware of additional artifacts arising from choices in the implementation of each method. 

The dependence of Bryan's MEM on the limited search space was among the central reasons for the development of the BR method. The advantageous form of the BR prior, which does not suffer from slow convergence in finding $\rho^{\rm MAP}$ in practice, allows one to carry out the needed optimization in the full $N_\mu$ dimensional solution space to $P[\rho|D,I]$ with reasonable computational cost. The proof from Ref.~\cite{Asakawa:2000tr} which also applies to the convex BR prior, guarantees that in the full search space a single unique Bayesian solution can be located if it exists. 

In \cref{sec:handson} we will take a look at hands-on examples of using the BR method to extract spectral functions and estimating their reliability.

\subsection{Two lattice QCD uncertainty challenges}

Spectral function reconstruction studies from lattice QCD have encountered two major challenges in the past.

The first one is related to the number of available input data points, which, compared to simulations in e.g. condensed matter physics is relatively small, of the order ${\cal}O(10-100)$. Especially when analyzing datasets at the lower end of this range, the sparsity of the $D_i$'s along Euclidean time $\tau$ often translates into ringing artefacts. Due to the restricted search space of Bryan's MEM, this phenomenon may be hidden, while the global extremum of the MEM posterior $\rho^{\rm MAP}_{\rm MEM}$, as well as the BR method MAP estimate $\rho^{\rm MAP}_{\rm BR}$ do show ringing. Since ringing leads to spectral functions with a too large arc length compared to the true spectral function one can treat this artifact by combining either the MEM or the BR prior with the arc-length penalty regulator discussed in \cref{sec:bayedspecrec}. The additional hyperparameters associated with this penalty term can be estimated using realistic mock data, as shown e.g. in Ref.~\cite{Kim:2018yhk}. The benefit of this genuine Bayesian approach is that the mechanism by which ringing is suppressed is made explicit and is not hidden in a particular choice of basis function.

The second challenge affects predominantly spectral reconstructions at finite temperature, in particular their comparability at different temperatures. In lattice QCD, temperature is encoded in the length of the imaginary time axis. I.e. simulations at lower temperature have access to a larger $\tau$ regime, as those at higher temperature. Since the available Euclidean time range affects the resolution capabilities of any spectral reconstruction it is important to calibrate one's results to a common baseline. I.e. one needs to establish how the accuracy of the reconstruction method changes as one increases temperature. Otherwise changes in the reconstructed spectral functions are attributed to physics, while they actually represent simply a degradation of the method's resolution. The concept of the \textit{reconstructed correlator} \cite{Datta:2003ww} is an important tool in this regard. Assume we have a correlator encoding a certain spectral function at temperature $T_1$ with $N_{\tau}^{\rm T_1}$ points. We can now ask: how would the correlator look like where the same spectral function is encoded at a higher temperature $T_2$, i.e. within a smaller Euclidean time window of $N_\tau^{\rm T_2}$ points. Since the underlying kernel relating spectral function and correlator is often temperature dependent, this question is not easily answered by just discarding imaginary time datapoints from the large $\tau$ region of the original correlator\footnote{In cases where the kernel is temperature independent, e.g. for lattice effective field theory correlators, discarding large $\tau$ datapoints is equivalent to computing the reconstructed kernel.}. Instead if one wishes to evaluate the corresponding higher temperature correlator Ref.~\cite{Ding:2012sp} showed that for the bosonic finite temperature kernel $K^{\rm T>0}(\mu,\tau)={\rm cosh}[\mu(\tau-\beta/2)]/{\rm sinh}[\mu\beta/2]$, relevant for studies of relativistic bosonic spectral functions, one has to form the following quantity
\begin{align}
    D_{\rm rec}(\tau,T_2|T_1)=\sum_{\tau'/a=\tau/a,\Delta \tau'/a=N_\tau^{\rm T_1}}^{N_\tau^{\rm T_1}-N_\tau^{\rm T_2}+\tau/a}D_{\rm lattice}(\tau'|T_1).
\end{align}
By carrying out a reconstruction based on the two correlators at different Euclidean extent $D_{\rm lattice}(\tau|T_1)$ and $D_{\rm lattice}(\tau|T_2)$ one will in general obtain two different spectral functions, even though the encoded spectrum is the same. Only when one compares the reconstruction based on $D_{\rm rec}(\tau,T_2|T_1)$ with that of $D_{\rm lattice}(\tau|T_2)$ is it possible to disentangle the genuine effects of a change in temperature from the one's induced by the reduction in access to Euclidean time. This reconstruction strategy has been first deployed for relativistic correlators in Ref.~\cite{Kelly:2018hsi}. A similar analysis in the context of non-relativistic spectral functions in Ref.~\cite{Kim:2018yhk} showed that the temperature effect of a negative mass shift for in-medium hadrons was only observable, if the changes in resolution of the reconstruction had been taken into account.

\section{Hands-on spectral reconstruction with the BR method}
\label{sec:handson}

This publication is accompanied by two open-source codes. The first \cite{BRMAP}, written in the \texttt{C/C++} language, implements the BR method (and the MEM) in its traditional form to compute MAP estimates with arbitrary precision arithmetic. The second \cite{BRMCStan}, written in the Python language uses standard double precision arithmetic and utilizes the modern \texttt{MCStan} Monte-Carlo sampler to evaluate the full BR posterior.

\subsection{BR MAP implementation in \texttt{C/C++}}

The BR MAP code deploys arbitrary precision arithmetics, based on the \texttt{GMP} \cite{Granlund12} and \texttt{MPFR} \cite{10.1145/1236463.1236468} libraries, which offers numerical stability for systems where exponential kernels are evaluated over large frequency ranges. A run-script called \texttt{BAYES.scr} is provided in which all parameters of the code can be specified. 

The kernel for a reconstruction task is apriori known and depends on the system in question. The BR MAP code implements three common types encountered in the context of lattice QCD (see parameter \texttt{KERNELTYPE}). Both zero temperature kernel $K^{\rm T=0}(\mu,\tau)={\rm exp}[-\mu\tau]$, and the naive finite temperature kernel for bosonic correlators $K^{\rm T>0}(\mu,\tau)={\rm cosh}[\mu(\tau-\beta/2)]/{\rm sinh}[\mu\beta/2]$ are available. Here $\beta$ refers to the extend of the imaginary time axis. The third option is the regularized finite temperature kernel $K^{\rm T>0}_{\rm reg}(\mu,\tau)=\frac{\beta}{2\pi}{\rm atan}[\mu]K^{\rm T>0}(\mu,\tau)$ suggested in Ref.~\cite{Ding:2012sp} (see also \cite{Aarts:2007wj,Ding:2009ie}). It lifts the divergence of the kernel at $\mu=0$, which is related to the antisymmetry of bosonic spectral functions at $T>0$. Note that when redefining the kernel, one also redefines the spectral function to reconstruct and thus an appropriately modified default model must be supplied.

Next, the discretization of the frequency interval $\mu$ needs to be decided on (see parameters \texttt{WMIN} and \texttt{WMAX}). When relativistic lattice QCD correlators are investigated, the lattice cutoff $\pm\sqrt{3}\frac{\pi}{a}$ provides a reliable estimate up to where spectral structures will be present. It is often a good crosscheck to use a larger range of frequencies beyond where the input data can provide constraining information, in order to see that the reconstructed spectral function in that regime is correctly given by the supplied default model. In case that lattice effective field theory correlators are investigated, the user has to keep in mind that their spectra may be populated beyond the naive lattice cutoff. In some cases the appropriate range can be estimated from an inspection of semi-analytically tractable free theory spectral functions. A rough guess for the UV cutoff can be obtained by fitting an exponential to the first few correlator points at small imaginary time $\tau$. Depending on the resolution required for the encoded spectral features, the number of frequency bins $N_\mu$ can be chosen via \texttt{NOMEGA}. If a very sharp peak feature is present, one can use the parameters \texttt{HPSTART}, \texttt{HPEND} and \texttt{HPNUM} to define a high resolution window along $\mu$ for which \texttt{HPNUM} of the \texttt{NOMEGA} points are used.

The number of points along the Euclidean time axis of the lattice simulation is specified by \texttt{NT} and its extend noted by \texttt{BETA}. Depending on the form of the kernel and the choices for $\beta$ and $\mu_{\rm max}$ the dynamic range of the kernel matrix may be large and one has to choose an appropriate precision \texttt{NUMPREC} for the arithmetic operations used.

For the analysis of lattice QCD correlators \texttt{FILEFORMAT} $4$ is most useful.  Each of the total \texttt{NUMCONF} measurements of a correlator is expected to be placed in individual files with a common name \texttt{DATANAME} (incl. directory information) and a counter as extension, which counts upward from \texttt{FOFFSET}. The format of each file is expected to contain two columns in \texttt{ASCII} format, the first denoting the Euclidean time step as integer and the second one the real-valued Euclidean correlator. Via \texttt{TMIN} and \texttt{TMAX} the user can specify which are the smallest and largest Euclidean times provided in each input data file, while \texttt{TUSEMIN} and \texttt{TUSEMAX} define which of these datapoints are used for the reconstruction.

In order to robustly estimate the statistical uncertainty of the input data, the code is able to perform an analysis of the autocorrelation among the different measurements. The value of \texttt{ACTHRESH} is used to decide to which threshold the normalized autocorrelation function \cite{montvay1994quantum} must have decayed, for us to consider subsequent measurements as uncorrelated. To test the quality of the estimated errors one can manually enlarge or shrink the assigned error values using the parameter \texttt{ERRADAPTION}.

As discussed in the previous section, a robust estimate of the statistical uncertainty of the spectral reconstruction can be obtained from a Jackknife analysis. The code implements this type of error estimate when the number of Jackknife blocks are set to a value larger than two in \texttt{JACKNUM}. The \texttt{NUMCONF} measurements are divided into consecutive blocks and in each iteration of the Jackknife a single block is remove when computing the mean of the correlator. If \texttt{JACKNUM} is set to zero a single reconstruction based on the full available statistics is carried out.

Once the data is specified, we have to select the default model. The default model can either be chosen to take on a simple functional form choosing values $1$ or $2$ for \texttt{PRIORMODEL}. The latter corresponds to a constant given by \texttt{MFAC}. The former leads to $m(\mu)=m_0/(\mu-\mu_{\rm min}+1)^{\rm power}$, where the power is set via the parameter \texttt{PRIORPOWER} and $m_0$ via \texttt{MFAC}. To supply more elaborate default models the user can set \texttt{PRIORMODEL} to $4$ and provide a file \texttt{prior.0} in the working directory of the code that contains two columns, the first with the frequencies $\mu$ and the second with the values of $m$. Note that we have already marginalized over the uncertainty of the default model using $P[\alpha]=1$ so that specifying $m$ suffices for the BR method.

In the present implementation of the BR method (\texttt{ALGORITHM} value $1$) the integration over $\alpha$ is implemented in a semi-analytic fashion, which is based on a large $S$ expansion. In practice this simply means that one must avoid to start the minimizer from the default model for which $S=0$. 

The original Ref.~\cite{Burnier:2013nla} conservatively stated that it is advantageous with regards to avoiding overfitting to instruct the minimizer to keep the values of the likelihood close to the number of provided datapoints. The code maintains this condition within a tolerance that is specified by a combination of the less than ideal named \texttt{ALPHAMIN} and \texttt{ANUM} parameters. The reconstruction will be performed \texttt{ANUM} times where in each of the iterations counted by \texttt{ANUM} the likelihood is constrained to fulfill $|L-N_{\rm data}|=(1/$\texttt{ALPHAMIN}$\times 10^{\rm ACNT})$.
 
The search for $\rho^{\rm MAP}_{\rm BR}$ is carried out internally using the LBFGS minimization algorithm \cite{10.3115/1118853.1118871}. It terminates when the step size of the minimizer falls below the threshold \texttt{MINTOL}. Note that for high precision arithmetic a correspondingly small threshold should be specified (e.g. for \texttt{NUMPREC}=128 \texttt{MINTOL}=$10^{-30}$ or for \texttt{NUMPREC}=256 \texttt{MINTOL}=$10^{-60}$).  The results of the minimizer are output into the folder \texttt{RESULTNAME} every $2000$ steps in files called \texttt{BAYES\_rhovalues\_A(ANUM-ACNT).dat} and the final result is found in the file \texttt{spec\_rec.dat}. The spectra are also collected in the file \texttt{PROB\_ESTIMATES\_FREQ.dat} in column $6$, where the frequencies are listed in column $4$. If the Jackknife analysis is selected then this file contains multiple spectra for each Jackknife subaverage counted by the value in column $8$. 

To speed up the convergence in case that very high precision data is supplied (i.e. when very sharp valleys exist in the likelihood) it is advantageous to carry out the reconstruction first with artificially enlarged errorbars via \texttt{ERRADAPTION}$>1$. The corresponding result in file \texttt{BAYES\_rhovalues\_A(ANUM).dat} if copied into the working directory of the code with the name \texttt{start.0} can be used as starting point for the next minimization with the actual errorbars, by selecting the value $2$ for the parameter \texttt{RESTARTPREV}.

The code, when compiled with the preprocessor macro \texttt{VERBOSITY} set to value one, will give ample output about each step of the reconstruction. It will output the frequency discretization, the values for the Euclidean times used, as well as show which data from each datafile has been read-in. In addition it presents the estimated autocorrelation and the eigenvalues of the covariance matrix, before outputting each step of the minimizer to the terminal. This comprehensive output allows the user to spot potential errors during data read-in and allows easy monitoring whether the minimizer is proceeding normally. The incorrect estimation of the covariance matrix due to autocorrelations is a common issue, which can prevent the minimizer to reach the target of minimizing the likelihood down to values close to the number of input data. Enlarging the errorbars until the likelihood reaches small enough values provides a first indication of how badly the covariance matrix is affected by autocorrelations. Another diagnosis step is to only consider the diagonal entries of the covariance matrix, which can be selected using the preprocessor macro \texttt{DIAGCORR} set to $1$.

\subsection{MEM MAP implementation in \texttt{C/C++}}

The provided \texttt{C/C++} code also allows to perform the MAP estimation based on the MEM prior using arbitrary precision arithmetic. By setting the parameter \texttt{ALGORIHM} to value $2$ once can choose Bryan's implementation, where the spectral function is parameterized via the SVD of the kernel matrix. The standard implementation uses as many SVD basis functions as input datapoints are provided. By varying the \texttt{SVDEXT} parameter the user may choose to include more or reduce the number of SVD basis function deployed. Alternatively by using the value $3$ the user can deploy the Fourier basis functions introduced in Ref.~\cite{Rothkopf:2012vv} and for value $4$ $\rho^{\rm MAP}_{\rm MEM}$ is searched for in the full $N_\mu$ dimensional search space. Due to the proof of uniqueness of the extremum, even searching in the full space is supposed to locate a single Bayesian answer $\rho^{\rm MAP}_{\rm MEM}$ to the inverse problem.

In the MEM, the common uncertainty parameter $\alpha$ for the default model $m_l$ is still part of the posterior and needs to be treated explicitly. To this end the MEM reconstruction is repeated \texttt{ANUM} times, scanning a range of $\alpha$ values between \texttt{ALPHAMIN} and \texttt{ALPHAMAX}. Since apriori the appropriate range of values is not known, the user is recommended to carry out reconstructions with artificially enlarged errorbars via \texttt{ERRADAPTION} that converge quickly and which allow to scan a large range between usually $\alpha\in[0;100]$.

The LBFGS minimizer will be used to find the point estimates $\rho^{\rm MAP}_\alpha$ for each fixed value of the hyperparameter and then according to Ref.~\cite{Asakawa:2000tr} estimate the probability distribution $P[\alpha|D,I]$ over which a weighted average is computed. The final result is then outputted in the file \texttt{spec\_rec.dat} in column $4$ with the frequencies located in column $3$. Intermediate steps of the minimizer are output to files \texttt{MEM\_rhovalues\_A(ACNT).dat}, where \texttt{ACNT} refers to the step along the alpha interval. In case of a Jackknife analysis all reconstructed spectra can be found in \texttt{PROB\_ESTIMATES\_FREQ.dat} in column $6$, where the frequencies are listed in column $4$.

Note that due to the functional form of the Shannon-Jaynes prior the convergence for spectral functions with large regions of vanishing values is often slow, which is why in practice the tolerance for convergence is chosen by \texttt{MINTOL} around $10^{-7}$.

Note that the estimation of the $\alpha$ probabilities involves the computation of eigenvalues of a product of the kernel with itself. In turn this step may require additional numerical precision via \texttt{NUMPREC} if an exponential kernel is used. If the precision is insufficient, the determination of the eigenvalues might fail and the final integrated spectral function will show \texttt{NAN} values, while intermediate results in \texttt{MEM\_rhovalues\_A(ACNT).dat} are well behaved. In that case rerunning the reconstruction with higher precision will remedy the issue.

\subsection{Full Monte-Carlo based BR method in Python}
\label{sec:BRMC}
In many circumstances the MAP point estimate of spectral functions already provides relevant information to answer questions about real-time physics from lattice QCD. However, as discussed in the previous section \cref{sec:uncertmap}, its full uncertainty budget may be challenging to estimate. It is therefore that I here discuss a modern implementation of the BR method, allowing for access to the posterior distribution via Monte-Carlo sampling.

The second code provided with this publication is a Python script based on the \texttt{MCStan} Monte-Carlo sampler library \cite{carpenter2017stan,standev2018stancore}. It uses the same parameters for the description of frequency and imaginary time as the \texttt{C/C++} code but works solely with double precision arithmetic. Since different kernels are easily re-implemented, the script contains as single example the zero temperature kernel $K^{\rm T=0}(\tau,\mu)$.

In order to sample from the posterior, we must define all the ingredients of our Bayesian model in the \texttt{MCStan} language. A simple model consists of three sections, \texttt{data}, \texttt{parameters} and the actual \texttt{model}. In \texttt{data} the different variables and vectors used in the evaluation of the model are specified. It contains e.g. the number of datapoints \texttt{sNt} and the number of frequency bins \texttt{sNw}. The decorrelated kernel is provided in a two-dimensional matrix datatype \texttt{Kernel}, while the decorrelated simulation data come in the form of a vector \texttt{D}. The eigenvalues of the covariance matrix enter via the vector \texttt{Uncertainty}. The values of the default model are stored in the vector \texttt{DefMod}. In the original BR method we would assume full ignorance of the uncertainty parameters $\alpha_l$ with $P[\alpha]=1$. Such improper priors may lead to inefficient sampling in \texttt{MCStan}, which is why in this example script a lognormal distribution is used. It draws $\alpha$ values from a range considered relevant in mock data tests. The user can always check self consistently whether the sampling range of $\alpha$'s was chosen appropriately by interrogating the marginalized posterior for $\alpha$ itself, making sure that its maximum lies well within the sampling range.

After selecting how many Markov-chains to initialize via \texttt{NChain} and how many steps in Monte Carlo time to proceed via \texttt{NSamples} the Monte-Carlo sampler of \texttt{MCStan} is executed using the \texttt{sample} command. MCStan automatically adds additional steps for thermalization of the Markov chain. Depending on how well localized the histrograms for each $\rho_l$ are, the number of samples must be adjusted. Since the BR prior is convex, initializing different chains in different regions of parameter space does not affect the outcome as long as enough samples are drawn.

We may then subsequently estimate the spectral function reconstruction from the posterior by inspecting the histograms for each parameter. Since in this case we have access to the full posterior distribution we can now answer not only what the most probable value for $\rho_l$ is but also compute its mean and median, giving us relevant insight about the skewness of the distribution of values. 

\subsection{Mock Data}

Both code packages contain two realistic mock-data test sets, which have been used in the past to benchmark the performance of Bayesian methods. They are based on the Euclidean Wilson loop computed in first order hard-thermal-loop perturbation theory, for which the temperature independent kernel $K(\tau,\mu)={\rm exp}[-\tau\mu]$ is appropriate. The correlator included here corresponds to the one computed at $T=631$MeV in Ref.~\cite{Burnier:2013fca} and which is evaluated at $r=0.066$fm, as well as $r=0.264$fm spatial extend. The continuum correlator is discretized with $32$ steps in Euclidean time. The underlying spectral functions are provided in the folder \texttt{MockSpectra} in separate files for comparison.

To stay as close to the scenario of a lattice simulation, based on the ideal correlator data, a set of 1000 individual datafiles is generated in the folder \texttt{MockData} in which the imaginary time data is distorted with Gaussian noise. The noise strength is set to give a constant $\Delta D/D=10^{-4}$ relative error on the mean when all samples are combined. The user is advised to skip both the first $D(0)$ and last datapoint $D(\tau_{\rm max})$ in the dataset, which are contaminated by unphysical artifacts related to the regularization of the Wilson loop computation.

The reader will find that this mock data provides a challenging setting for any reconstruction method, as it requires the reconstruction both of a well defined peak, as well as of a broad background structure. It therefore is well suited to test the resolution capabilities of reconstruction methods, as well as their propensity for ringing and over-damping artifacts. 

\begin{figure}
    \centering
    \includegraphics[scale=0.5]{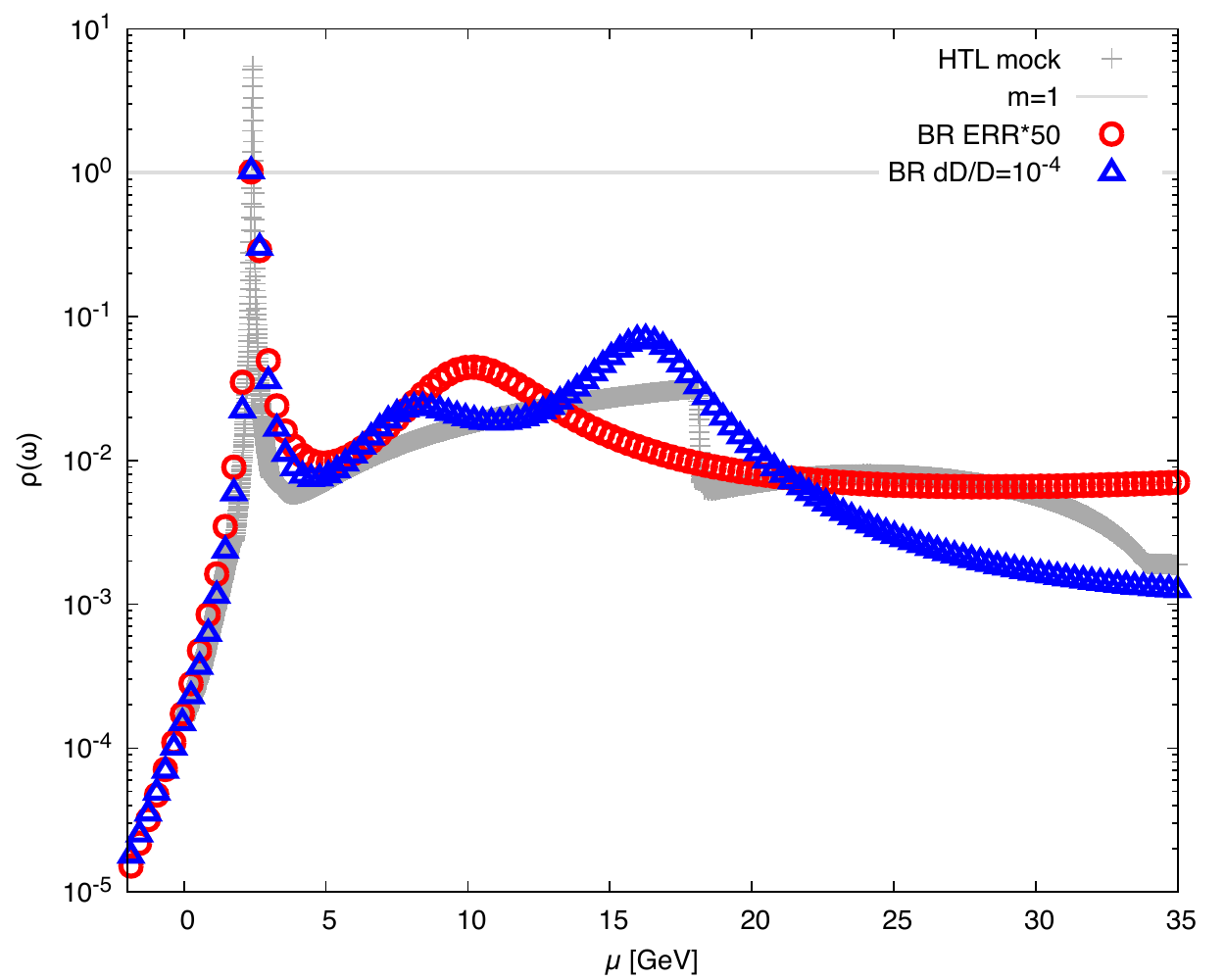}
    \includegraphics[scale=0.5]{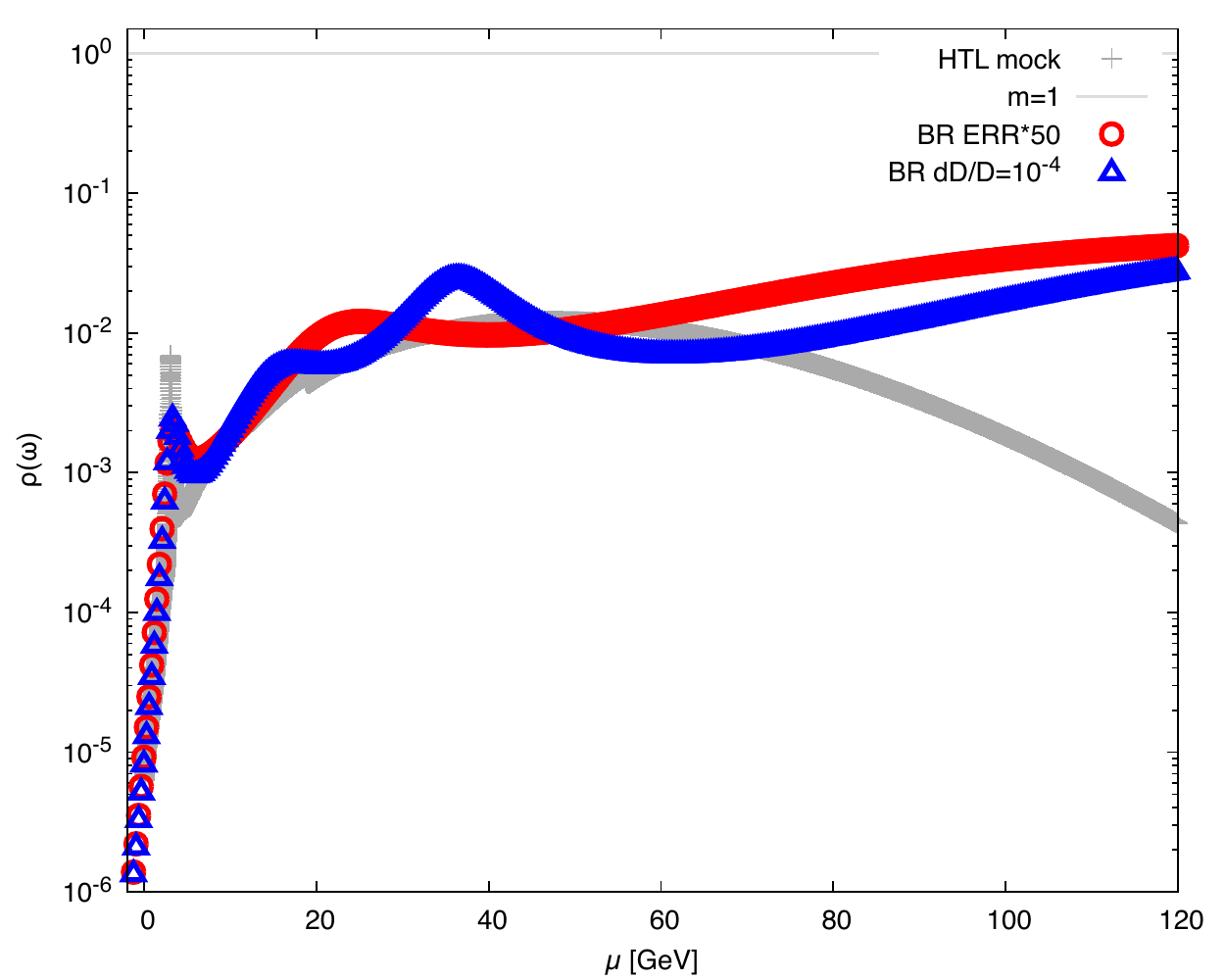}
    \caption{BR MAP reconstructions of the HTL Wilson loop spectral function (gray points) evaluated at $T=631$MeV and spatial separation distance $r=0.066$fm (top) and $r=0.264$fm (bottom). The reconstruction based on $N_\tau=32$ Euclidean data and a frequency range between $\mu a \in[-5,25]$ with $N_\omega=1000$ are shown as colored open symbols. The red data denotes the reconstruction based on the preconditioning \texttt{ERRADAPTION}$=50$ while the final result exploting the full $\Delta D/D=10^{-4}$ is given in blue.}
    \label{fig:HTLmockanalysis}
\end{figure}

For the \texttt{C/C++} implementation of the BR MAP estimation a set of example scripts are provided. The user can first execute e.g. \texttt{BAYESMOCK066\_precon.scr} to carry out a preconditioning run with enlarged errorbars. In a second step one provides the outcome of the preconditioning run as file \texttt{start.0} and executes \texttt{BAYESMOCK066.scr} to locate the global extremum of the BR prior. The outcome of these sample scripts is given for reference in \cref{fig:HTLmockanalysis} compared to the semi-analytically computed HTL spectral functions in \texttt{SpectrumWilsonLoopHTLR066.dat}.

\section{New insight from machine learning}

Over the past years interest in machine learning approaches to spectral function reconstruction has increased markedly (see also \cite{Boyda:2022nmh}). Several groups have put forward pioneering studies that explore how established machine learning strategies, such as supervised kernel ridge regression \cite{Offler:2021fmg,Spriggs:2021dsb}, artificial neural networks \cite{fournier2020artificial,Kades:2019wtd,Karpie:2019eiq,Chen:2021giw,Wang:2021cqw,Shi:2022yqw} or Gaussian processes \cite{Horak:2021syv} can be used to tackle the inverse problem in the context of extracting spectral functions from Euclidean lattice correlators. The machine learning mindset has already lead to new developments in the spectral reconstruction community, by providing new impulses to regularization of the ill-posed problem.

As a first step let us take look at how machine learning strategies incorporate the necessary prior knowledge to obtain a unique answer to the reconstruction task. While in the Bayesian approach this information enters explicitly through the prior probability and its hyperparameters, it does so in the machine-learning context in three separate ways: To train supervised reconstruction algorithms a training dataset needs to be provided, often consisting of pairs of correlators and information on the encoded spectral functions. Usually a limited selection of relevant structures is included in this training data set, which amounts to prior knowledge on the spectrum. Both supervised and unsupervised machine learning is build around the concept of a cost- or optimization functional, which contains information on the provided data. It most often also features regulator terms, which can be of similar form as those discussed in \cref{sec:bayedspecrec}. This in particular means that these regulators define the most probable values for the $\rho_l$'s in the absence of data and therefor take on a similar role as a Bayesian default model. The third entry point for prior knowledge lies in the choice of structures used to compose the machine learning model. In case that e.g. Gaussian processes are used, the choice of kernel of the common normal distribution for observed and unobserved data is based on prior knowledge, as is the selection of its hyperparameters. In case that neural networks are used, the number and structure of the deployed layers and activation functions similarly imprint additional prior information on the reconstructed spectral function, such as e.g. their positivity.

Direct applications of machine learning approaches developed in the context of image reconstruction to positive spectral function reconstruction have shown good performance on-par with Bayesian algorithms, such as the BR method or the MEM. 

Can we understand why machine learning so far has not outpaced Bayesian approaches? One potential answer lies in the information scarcity of the input correlators themselves. If there is no unused information present in the correlator also sophisticated machine learning cannot go beyond what Bayesian approaches utilize. As shown in recent mock-data tests in the context of finite temperature hadron spectral functions in Ref.~\cite{Kim:2018yhk}, increasing the number of available datapoints in imaginary time (i.e. going closer to the continuum limit) does not necessarily improve the reconstruction outcome significantly as the relevant information content about thermal physics does not increase. This is easily seen when considering the Matsubara frequency correlator. As one decreases the temporal lattice spacing, the range of accessible high lying Matsubara frequencies increases but their coarseness, given by the inverse temperature of the system remains the same. It turns out that the relevant thermal physics is often hidden in the range between the first and second Matsubara frequency and the correlator at higher frequencies already coincides within errors with the zero temperature correlator. 

This information scarcity dilemma asks us to provide our reconstruction algorithms with more QCD specific prior information. So far the Bayesian priors have focused on very generic properties, such as positivity and smoothness. It is here that machine learning can and already has provided new impulses to the community. 

One promising approach is to use neural networks as parameterization of spectral functions or parton distribution functions. First introduced in the context of PDFs in Ref.~\cite{Karpie:2019eiq} and recently applied to the study of finite temperature spectra in Ref.~\cite{Shi:2022yqw} this approach allows to infuse the reconstruction with additional information about the analytic properties of $\rho$. Traditionally one would choose a specific parameterization apriori such as rational functions (Pad\'e) or SVD basis functions (Bryan) and vary their parameters. The more versatile NN approach, thanks to the universal approximation theorem, allows us instead to explore different types of basis functions and assign an uncertainty to each choice.

The concept of learning can also be brought to the prior probability or regulator itself. Instead of constructing a regulator based on generic axioms, one may consider it as a neural network mapping the parameters $\rho_l$ to a single penalty value $P[\rho|I]$. Training an optimal regulator within a Bayesian setting, based e.g. on realistic mock data, promises to capture more QCD specific properties than what is currently encoded in the BR or MEM.  Exploring this path is work in progress.

\section{Summary and Conclusion}

Progress in modern high-energy nuclear physics depends on first-principle knowledge of QCD dynamics, be it in the form of transport properties of quarks and gluons at high temperatures or the phase-space distributions of partons inside nucleons at low temperatures. Lattice QCD offers non-perturbative access to these quantities but due to its formulation in imaginary time, hides them behind an ill-posed inverse problem. The inverse problem is most succinctly stated in terms of a spectral decomposition, where the Euclidean correlator accessible on the lattice is expressed as integral over a spectral function multiplied by an analytic kernel. The real-time information of interest can often be read-off directly from the structures occurring in the spectral function. The determination of PDFs from the hadronic tensor and via pseudo PDFs can be formulated in terms of a similar inversion problem.

Bayesian inference provides a versatile tool set for the reconstruction of spectral functions. It gives meaning to the ill-posed inverse problem by incorporating relevant domain knowledge with an associated uncertainty budget through the prior probability distribution. Evaluating the posterior distribution, defined through Bayes theorem, gives access to the most probable values of the spectral function based on simulation data and prior knowledge. In addition it also encodes the full uncertainty budget through its spread. Traditionally predominantly MAP point estimates were computed due to lower computational cost of the corresponding optimization problem, compared to full Monte-Carlo sampling of the posterior. In that case information about the uncertainty budget is hidden from the user and it must be estimated manually. Several relevant challenges for uncertainty estimation in the lattice QCD context were discussed, including the problem of ringing and those related to comparing reconstructions based on different Euclidean time extents.

A brief user guide described how to run two open access codes accompanying this publication. One focuses on the determination of MAP point estimates based on the BR and MEM prior. The other utilizes a modern Monte-Carlo library to sample from the full BR posterior. 

Last but not least a brief look is taken at machine learning approaches to spectral function reconstruction. The need for providing prior information is discussed and a common challenge among all reconstruction approaches, information scarcity in the input data, is pointed out. Two venues for combining the machine-learning viewpoint with the Bayesian strategy are touched upon.

With the concrete conceptual and technical discussions contained in this publication, the reader is equipped with a solid basis to carry out Bayesian spectral reconstructions. The provided open-access source codes offer a quick entry into the research field and can be modified according to different needs in regards to kernels arising in different lattice QCD studies.

\section*{Acknowledgements}
The author gladly acknowledges support by the Research Council of Norway under the FRIPRO Young Research Talent grant 286883. Some of the spectral function reconstruction code has been developed in the context of the project NN9578K-QCDrtX "Real-time dynamics of nuclear matter under extreme conditions" funded by UNINETT Sigma2 - the National Infrastructure for High Performance Computing and Data Storage in Norway.  

%%%%%%%%%%%%%%%%%%%%%%%%%%%%%%%%%%%%%%%%%
%%                                     									       %%
%%  So called Backmatter part starts here with acknowledgements		       %%
%%  funding information, and the bibliography.						       %%
%%                                     									       %%
%%%%%%%%%%%%%%%%%%%%%%%%%%%%%%%%%%%%%%%%%

\FloatBarrier

\begin{backmatter}

\section*{Competing interests}
  The author declares that he has no competing interests.

\section*{Author's contributions}
    \begin{itemize}
        \item A. Rothkopf: conception, literature study, code development, writing, editing.
    \end{itemize}

%%%%%%%%%%%%%%%%%%%%%%%%%%%%%%%%%%%%%%%%%
%%                                     									       %%
%%  Bibliography part starts here								       %%
%%                                     									       %%
%%%%%%%%%%%%%%%%%%%%%%%%%%%%%%%%%%%%%%%%%

\bibliographystyle{stavanger-mathphys}

%%%%%%%%%%%%%%%%%%%%%%%%%%%%%%%%%%%%%%%%%
%%                                     									       %%
%%  Specify your BibTeX bibliography file here or manually insert references  %%
%%                                     									       %%
%%%%%%%%%%%%%%%%%%%%%%%%%%%%%%%%%%%%%%%%%

\bibliography{references}

% or include bibliography directly:
% \begin{thebibliography}
% \bibitem{b1}
% \end{thebibliography}

\end{backmatter}

%%%%%%%%%%%%%%%%%%%%%%%%%%%%%%%%%%%%%%%%%
%%                                     									       %%
%%  End of the document										       %%
%%                                     									       %%
%%%%%%%%%%%%%%%%%%%%%%%%%%%%%%%%%%%%%%%%%

\end{document}